\documentclass[aps,superscriptaddress, showpacs,preprintnumbers, superscriptaddress, nofootinbibt,twocolumn]{revtex4}
\usepackage{amsmath}
\usepackage{eurosym}
\usepackage{amssymb}
\usepackage{graphicx}
\usepackage{color}

\def\cL{\mathcal{L}}
\def\be{\begin{equation}}
\def\ee{\end{equation}}
\def\bea{\begin{eqnarray}}
\def\eea{\end{eqnarray}}

\def\nn{\nonumber \\}

\begin{document}

\title{Cosmological implications of modified gravity induced by quantum metric fluctuations}

\author{Xing Liu}
\email{309166138@qq.com}
\affiliation{ School
of Physics,
Sun Yat-Sen University, Guangzhou 510275, P. R. of China}
\affiliation{Yat Sen School, Sun Yat-Sen University, Guangzhou 510275, P. R. China}
\author{Tiberiu Harko}
\email{t.harko@ucl.ac.uk}
\affiliation{Department of Physics, Babes-Bolyai University, Kogalniceanu Street,
Cluj-Napoca 400084, Romania,}
\affiliation{Department of Mathematics, University College London, Gower Street, London
WC1E 6BT, United Kingdom}
\author{Shi-Dong Liang}
\email{stslsd@mail.sysu.edu.cn}
\affiliation{State Key Laboratory of Optoelectronic Material and Technology, and
Guangdong Province Key Laboratory of Display Material and Technology, School
of Physics,\\
Sun Yat-Sen University, Guangzhou 510275, P. R. China}
\date{\today }

\begin{abstract}
We investigate the cosmological implications of modified gravities induced by the quantum fluctuations of the gravitational metric. If the metric can be decomposed as the sum of the classical and of a fluctuating part, of quantum origin, then the corresponding Einstein quantum gravity generates at the classical level modified gravity models with a nonminimal coupling between geometry and matter. As a first step in our study, after assuming that the expectation value of the quantum correction can be generally expressed in terms of an arbitrary second order tensor constructed from the metric and from the thermodynamic quantities characterizing the matter content of the Universe, we derive the (classical) gravitational field equations in their general form. We analyze in detail the cosmological models obtained by assuming that the quantum correction tensor is given by the coupling of a scalar field and of a scalar function to the metric tensor, and by a term proportional to the matter energy-momentum tensor. For each considered model we obtain the gravitational field equations, and the generalized Friedmann equations for the case of a flat homogeneous and isotropic geometry. In some of these models the divergence of the matter energy-momentum tensor is non-zero, indicating a process of matter creation, which corresponds to an irreversible energy flow from the gravitational field to the matter fluid, and which is direct consequence of the nonminimal curvature-matter coupling. The cosmological evolution equations of these modified gravity models induced by the quantum fluctuations of the metric are investigated in detail by using both analytical and numerical methods, and it is shown that a large variety of cosmological models can be constructed, which, depending on the numerical values of the model parameters, can exhibit both accelerating and decelerating behaviors.
\end{abstract}


\maketitle

\section{Introduction}

Modified gravity theories may provide an attractive alternative to the standard explanations of the present day observations that have shaken the well-established foundations of the theoretical physics.  Astronomical observations have confirmed that our Universe does not according to standard general relativity, as derived from the Hilbert-Einstein action, $S=\int{\left(-R/2\kappa ^2+L_m\right)\sqrt{-g}d^4x}$, where $R$ is the Ricci scalar, $\kappa $ is the gravitational coupling constant, and $L_m$ is the matter Lagrangian, respectively. Extremely successful on the Solar System scale, somehow unexpectedly, general relativity faces, on a fundamental theoretical level, two important challenges, the dark energy and the dark matter problems, respectively. Several high precision astronomical observations, with the initial goal of improving the numerical values of the basic cosmological parameters  by using the properties of the distant
Type Ia Supernovae, have provided the  result that the Universe underwent recently a transition to an accelerating, de Sitter type phase \cite{1n,2n,3n,4n,acc}. The necessity of explaining the late time acceleration lead to the formulation of a new paradigm in theoretical physics and cosmology, which postulates that the explanation of the late time acceleration is the existence of a mysterious component of the Universe, called
dark energy (DE), which can describe  the
late time dynamics of the Universe \cite{PeRa03,Pa03}, and can explain all the observed features of the recent (and future) cosmological evolution. However, in order to close the matter-energy balance of the Universe, a second, and equally mysterious component, called Dark Matter, is required. Dark Matter, assumed to be a non-baryonic and non-relativistic (cold) component of the Universe, is necessary for explaining the dynamics of the hydrogen clouds rotating around galaxies, and having flat rotation curves, as well as
the virial mass discrepancy in clusters of galaxies \cite{dm1,dm2}. The direct detection/observation of the dark matter is extremely difficult due to the fact that it interacts only gravitationally with the baryonic matter. After many decades of intensive observational and experimental efforts there is no direct evidence on the particle nature of the dark matter.

One of the best theoretical descriptions that fits almost perfectly the observational data is based on the simplest theoretical extension of general relativity, which includes in the gravitational field equations the cosmological constant $\Lambda $ \cite{Wein,Wein1}. Based on this theoretical formalism the basic paradigm of modern cosmology has been formulated as the $\Lambda $CDM-$\Lambda $ Cold Dark Matter Model, in which the dark energy is nothing but the simple constant introduced almost one hundred years ago by Einstein. Even that the $\Lambda $CDM model fits the data well, it raises some fundamental theoretical questions about the possibility of explaining it. There is no theoretical explanation for the physical/geometrical nature of $\Lambda $, and, moreover, general relativity cannot give any hints on why it is so small, and why it is so fine tuned \cite{Wein,Wein1}. Therefore, the possibility that dark energy can be explained as an intrinsic property of a {\it generalized gravity theory}, going beyond general relativity, and its Hilbert-Einstein gravitational action, cannot be rejected a priori. In this context a large number of modified gravity models, all trying to extend and generalize the standard Einsteinian theory of gravity, have been proposed. Historically, in going beyond Einstein gravity, the first, and most natural step, was to extend {\it the geometric part of the Hilbert-Einstein action}. One of the first attempts in this direction  is  represented by the $f(R)$ gravity theory, in
which the gravitational action is generalized to be an arbitrary function of the Ricci scalar,  so that $S = \frac{1}{2\kappa ^2}\int {f(R) \sqrt{-g}d^4 x} +\int {L_m \sqrt{-g}d^4 x}$ \cite{Bu701,re4,Bu702,Bu703,re1,Bu704,Fel}. However, this as well as several other modifications of the Hilbert-Einstein action focussed  only on the geometric part of the gravitational action, by explicitly postulating that the matter Lagrangian plays {\it a subordinate and passive role} only, as compared to the geometry \cite{Mat}. From a technical point of view such an approach implies a  minimal coupling between matter and  geometry. But a fundamental theoretical principle, forbidding an arbitrary coupling between matter and geometry, has not been formulated yet, and perhaps it may simply not exist. On the other hand, if general matter-geometry  couplings are introduced, a large number of  theoretical gravitational models, with extremely interesting physical and cosmological properties, can be easily constructed.

The first of the modified gravity theory with arbitrary geometry-matter coupling was the $f\left( R,L_{m}\right) $  modified gravity theory \cite{fL1,fL2,fL3,fL4}, in which the total gravitational action takes the form $S = \frac{1}{2\kappa ^2}\int {f\left(R,L_m\right) \sqrt{-g}d^4 x} $. In this kind of theories matter is essentially indistinguishable from geometry, and plays an active role in generating the geometrical properties of the space-time.  A different geometry-matter coupling is introduced in the
$f(R,T)$ \cite{fT1,fT2} gravity theory, where matter and geometry are coupled via $T$, the trace of the energy-momentum tensor. The gravitational action of the $f(R,T)$ theory is given by $S=\int{\left[f(R,T)/2\kappa ^2+L_m\right]\sqrt{-g}d^4x}$. A recent review of the generalized $f\left(R,L_{m}\right)$ and $f(R,T)$ type
gravitational theories with non-minimal curvature-matter couplings can be found in \cite{Revn}. Several other gravitational theories involving geometry-matter couplings have also been proposed, and extensively studied, like, for example, the
Weyl-Cartan-Weitzenb\"{o}ck (WCW) gravity theory \cite{WCW}, hybrid metric-Palatini $%
f(R,\mathcal{R})$ gravity \cite{HM1,HM2,Revn1}, where $\mathcal{R}$ is the Ricci scalar
formed from a connection independent of the metric, $%
f\left(R,T,R_{\mu \nu }T^{\mu \nu }\right)$ gravity theory, where $R_{\mu \nu
} $ is the Ricci tensor, and $T_{\mu \nu }$ the matter energy-momentum
tensor, respectively \cite{Har4,Odin}, or $f(\tilde{T},\mathcal{T})$ gravity \cite{HT}, in which a
coupling between  the torsion scalar $\tilde{T}$, essentially a geometric quantity,  and  the trace $T$ of the matter
energy-momentum tensor is introduced. Gravitational models with higher derivative matter fields were investigated in \cite{HDM}.

One of the interesting (and intriguing) properties of the gravitational theories with geometry-matter coupling is the non-conservation of the matter energy-momentum tensor, whose four-divergence is usually different of zero, $\nabla _{\mu}T^{\mu \nu}\neq 0$. This property can be interpreted, from a thermodynamic  point of view,  by using the formalism of open thermodynamic systems \cite{fT2}. Hence one can assume that the generalized conservation equations in these gravitational theories describe {\it irreversible matter creation processes}. Thus the non-conservation of the energy-momentum tensor describes an irreversible energy flow from the gravitational field to the newly created matter constituents, with the second law of thermodynamics requiring that space-time transforms into matter. In \cite{fT2} the equivalent particle number creation rates, the creation pressure and the entropy production rates were obtained for both $f\left(R,L_m \right)$  and $f(R,T)$  gravity theories. The temperature evolution laws of the newly created particles was also obtained, and studied. Due to the non-conservation of the energy-momentum tensor, which is a direct consequence of the geometry--matter coupling, during the cosmological evolution of the Universe a large amount of comoving entropy could be also produced.

The prediction of the production of particles from the cosmological vacuum is one of the remarkable results of the quantum field theory in curved space-times \cite{P1,Z1,P2,Full,P3}. Particles creation processes are supposed to play a fundamental role in the quantum field theoretical approaches to gravity, where they naturally appear. It is a standard result of quantum field theory in curved spacetimes that quanta of the minimally-coupled scalar field are created in the expanding Friedmann-Robertson-Walker Universe \cite{P3}. Therefore, the presence of particle creation processes in both quantum theories of gravity and modified gravity theories with geometry-matter coupling may suggest that a deep connection between these two, apparently very different physical theories, may exist. And, interestingly enough, such a connection has been found in \cite{re11}, where it was pointed out that by using a nonperturbative approach for the quantization of the metric, proposed in \cite{re8,re9,re10},  a particular type of $f(R,T)$ gravity, with Lagrangian given by $L=\left[(1-\alpha )R/2\kappa ^2+\left(L_m-\alpha T/2\right)\right]\sqrt{-g}$, where $\alpha $ is a constant, naturally emerges as a result of the quantum fluctuations of the metric. This result suggests  that an equivalent microscopic quantum description of the matter creation processes in $f(R,T)$ or $f\left(R,L_m\right)$ gravity is possible, and such a description could  shed some light on the physical mechanisms leading to particle generation via gravity and matter geometry coupling. Such mechanisms do indeed exist, and they can be understood, at least qualitatively, in the framework of some quantum/semiclassical gravity models.

 It is the goal of the present paper to further investigate the cosmological implications of modified gravities induced by the quantum fluctuations of the gravitational metric, as initiated in \cite{re11,re8,re9,re10}. As a starting point we assume that a general quantum  metric can be decomposed as the sum of the classical and of a fluctuating part, the latter being of quantum (or stochastic) origin. If such a decomposition is possible,  the corresponding Einstein quantum gravity generates at the classical level modified gravity models with a nonminimal interaction between geometry and matter, as previously considered in \cite{fL1, fT1,Har4}. After assuming that the expectation value of the quantum correction can be generally expressed in terms of an arbitrary second order tensor $K_{\mu \nu}$, which can be constructed from the metric and from the thermodynamic quantities characterizing the matter content of the Universe, we derive from the first order quantum gravitational action the (classical) gravitational field equations in their general form. We analyze in detail the cosmological models obtained from the quantum fluctuations of the metric in tqo cases. First we assume that the quantum correction tensor $K_{\mu \nu}$ is given by the coupling of a scalar field and of a scalar function to the metric tensor, respectively. As a second case we consider that $K_{\mu \nu}$ is given by a term proportional to the matter energy-momentum tensor. The first choice gives a particular version of the $f(R,T)$ gravity model \cite{fT1}, while the second choice corresponds to specific case of the modified gravity theory of the form $f\left(R,T,R_{\mu \nu}T^{\mu \nu},T_{\mu \nu}T^{\mu \nu}\right)$ \cite{Har4}. For each considered model we obtain the gravitational field equations, and the generalized Friedmann equations for the case of a flat homogeneous and isotropic geometry. In some of these models the divergence of the matter energy-momentum tensor is non-zero, indicating a process of matter creation. From a physical point of view a non-zero divergence of the energy-momentum tensor can be interpreted as corresponding to an irreversible energy flow from the gravitational field to the matter fluid. Such an irreversible thermodynamic process is the direct consequence of the nonminimal curvature-matter coupling, induced in the present case by the quantum fluctuations of the metric \cite{fT2, creat, Pavon}. The cosmological evolution equations of these modified gravity models induced by the quantum fluctuations of the metric are investigated in detail by using both analytical and numerical methods. As a result of this analysis we show that a large variety of cosmological models can be constructed. Depending on the numerical values of the model parameters, these cosmological models can exhibit both late time accelerating, or decelerating behaviors.

 The present paper is organized as follows. In Section~\ref{sect2} we derive the general set of field equations induced by the quantum fluctuations of the metric. The relation between this approach and the standard semi-classical formulation of quantum gravity is also briefly discussed. In Section~\ref{sect3} we investigate in detail the cosmological implications of the quantum fluctuations induced modified gravity models with the fluctuation tensor proportional to the metric. Two cases are considered, in which the fluctuation couples to the metric via a scalar field, and a scalar function, respectively. Modified gravity models induced by quantum metric fluctuations proportional to the energy-momentum tensor are investigate in Section~\ref{sect4}. Finally, we discuss and conclude our results in Section~\ref{sect5}. The details of the derivation of the gravitational field equations for an arbitrary metric fluctuation tensor and for a fluctuation tensor proportional to the matter energy-momentum tensor are presented in Appendices~\ref{App1} and \ref{App2}, respectively. In the present paper we use a system of units with $c=1$.

\section{Modified gravity from quantum metric fluctuations}\label{sect2}

In the present Section we will briefly review the fluctuating metric approach to quantum gravity, we will discuss its relation with standard semiclassical gravity, and we will point out the quantum mechanical origins of the modified gravity models with geometry-matter coupling. Moreover, we derive the general set of field equations induced by the quantum fluctuations of the metric for an arbitrary form of the tensor $K_{\mu \nu}$.

\subsection{Modified gravity as the semiclassical approximation of quantum gravity}

In the standard quantum mechanics physical (or geometrical) quantities must be represented by operators. Therefore, in a full non-perturbative quantum approach the Einstein gravitational field equations must take an operator form, given by \cite{re8,re9,re10}
\begin{eqnarray}
\hat{R}_{\mu\nu}-\frac{1}{2}R\hat{g}_{\mu\nu}=\frac{8\pi G}{c^4}\hat{T}_{\mu\nu}.
\end{eqnarray}
As shown in \cite{re8,re9,re10}, in order to extract meaningful physical information from the Einstein operator equations one must average it over all possible products of the metric operators $\hat {g}\left(x_1\right)...\hat{g}\left(x_n\right)$, and thus to solve the infinite set of equations
\begin{eqnarray*}
<Q|\hat{g}(x_1)\hat{G}_{\mu\nu}|Q>&=&<Q|\hat{g}(x_1) \hat{T}_{\mu\nu}|Q>,\\
<Q|\hat{g}(x_1)\hat{g}(x_2) \hat{G}_{\mu\nu}|Q>&=&<Q|\hat{g}(x_1)\hat{g}(x_2) \hat{T}_{\mu\nu}|Q>,\\
\dots&=&\dots,
\end{eqnarray*}
for the Green functions $\hat{G}_{\mu\nu}$. In the above equations \(|Q>\) is quantum state that might not be the ordinary vacuum state. These equations cannot be solved analytically, and hence we have to use some approximations \cite{re8}-\cite{re10}. In \cite{re8} it was suggested to decompose the metric operator into the sum of an average metric $g_{\mu \nu}$, and a fluctuating part $\delta\hat{g}_{\mu\nu}$, according to
\begin{eqnarray}\label{c1}
\hat{g}_{\mu\nu}=g_{\mu\nu}+\delta\hat{g}_{\mu\nu}.
\end{eqnarray}

Assuming that
\be\label{c2}
<\delta \hat{g}_{\mu\nu}>=K_{\mu\nu}\ne 0,
\ee
where $K_{\mu\nu}$ is a classical tensor quantity,  and ignoring higher order fluctuations, the gravitational Lagrangian will be modified into \cite{re8}
\begin{eqnarray}\label{2}
\mathcal{L}&=&\mathcal{L}_g\left(\hat{g}_{\mu\nu}\right)+
\sqrt{-g}\mathcal{L}_m\left(\hat{g}_{\mu\nu}\right)\approx \mathcal{L}_g+\frac{\delta\mathcal{L}_g}{\delta g^{\mu\nu}}\delta\hat{g}^{\mu\nu}+\nonumber\\
&&\sqrt{-g}\mathcal{L}_m+
\frac{\delta(\sqrt{-g}\mathcal{L}_m)}{\delta g^{\mu\nu}}\delta\hat{g}^{\mu\nu}=-\frac{1}{2\kappa ^2}\sqrt{-g}\times \nonumber\\ &&\left(R+G_{\mu\nu}\delta\hat{g}^{\mu\nu}\right)+
\sqrt{-g}\Bigg(\mathcal{L}_m+\frac{1}{2}T_{\mu\nu}\delta\hat{g}^{\mu\nu}\Bigg),
\end{eqnarray}
where \(\kappa ^2=8\pi G/c^4\), and where we have defined the matter energy-momentum tensor as
\begin{eqnarray}
T_{_\mu\nu}=\frac{2}{\sqrt{-g}}\frac{\delta \left(\sqrt{-g}\mathcal{L}_m\right)}{\delta g^{\mu\nu}}.
\end{eqnarray}

Even that the above formalism starts from a full quantum approach of gravity, after performing the decomposition of the metric we are still considering semiclassical theories. In this paper we will consider several functional forms  of the so called quantum perturbation tensor \(K^{\mu\nu}\), we will obtain the field equations of the corresponding gravity theory, and we will investigate their cosmological implications, respectively.

The gravitational field equations corresponding to the first order corrected quantum Lagrangian (\ref{2}) are given, in a general form,  by
\bea\label{genfe}
G_{\mu\nu}=\kappa^2T_{\mu\nu}-
\Bigg\{\frac{1}{2}g_{\mu\nu}G_{\alpha\beta}K^{\alpha\beta}+\frac{1}{2}\times\nn
\Bigg(\Box K_{\mu\nu}+\nabla_{\alpha}\nabla_{\beta}K^{\alpha\beta}g_{\mu\nu}-
\nabla_{\alpha}\nabla_{(\mu}K_{\nu)}^\alpha\Bigg)+\nn
\gamma^{\alpha\beta}_{\mu\nu}R_{\alpha\beta}-\frac{1}{2}\Bigg[RK_{\mu\nu}+KR_{\mu\nu}+\gamma^{\alpha\beta}_{\mu\nu}(R g_{\alpha\beta})+\nn
\nabla_\mu\nabla_\nu K+g_{\mu\nu}\Box K\Bigg]\Bigg\}+\kappa^2\Bigg\{-\frac{1}{2}g_{\mu\nu}T_{\alpha\beta}K^{\alpha\beta}+\nn
\Bigg[\gamma^{\alpha\beta}_{\mu\nu}T_{\alpha\beta}+K^{\alpha\beta}\Bigg(2\frac{\delta^2 \mathcal{L}_m}{\delta g^{\mu\nu}\delta g^{\alpha\beta}}-\frac{1}{2}g_{\alpha\beta}T_{\mu\nu}\nn
-\frac{1}{2}g_{\mu\nu}g_{\alpha\beta}\mathcal{L}_m-\mathcal{L}_m\frac{\delta{g_{\alpha\beta}}}{\delta g^{\mu\nu}}\Bigg)\Bigg]\Bigg\},
\eea
 where \(K=g_{\mu\nu}K^{\mu\nu}\)and \(A_{\alpha\beta}\delta K^{\alpha\beta}=\delta g^{\mu\nu}(\gamma^{\alpha\beta}_{\mu\nu}A_{\alpha\beta})\). Here \(A_{\alpha\beta}\) is either $R_{\alpha \beta}$ or $T_{\alpha \beta }$, and \(\gamma^{\alpha\beta}_{\mu\nu}\) is an algebraic tensor, an operator, or the combination of them. The detailed derivation of Eqs.~(\ref{genfe}) is presented in Appendix~\ref{App1}.

 It is interesting to compare the formalism based on the quantum metric fluctuation proposal to the standard semiclassical gravity approach, which also leads to particle production via the non-conservation of the matter energy-momentum tensor \cite{Pavon}. Semiclassical gravity is constructed from the basic assumption that the gravitational field is still {\it classical}, while the {\it classical matter (bosonic) fields} $\phi $ are taken as {\it quantized}. The direct coupling of the  quantized matter fields to the classical gravitational fields is performed via the replacement of the quantum energy momentum tensor $\hat{T}_{\mu \nu}$ by its expectation value $\left <\hat{T}_{\mu \nu}\right >$, obtained with respect to some quantum state $\Psi $. Therefore  the effective semiclassical Einstein equation can be {\it postulated} as \cite{Carl},
\be\label{gr1}
R_{\mu \nu}-\frac{1}{2}g_{\mu \nu}R=\frac{8\pi G}{c^4}\left<\Psi \right |\hat{T}_{\mu \nu}\left |\Psi \right>.
\ee

From Eq.~(\ref{gr1})  it follows immediately that the classical energy-momentum tensor $T_{\mu \nu}$  of the gravitating system is obtained from its quantum counterpart through the definition  $\left<\Psi \right |\hat{T}_{\mu \nu}\left |\Psi \right>=T_{\mu \nu}$.
The semiclassical Einstein equations ~(\ref{gr1}) can be derived from the variational principle \cite{Kibble}
\be\label{quantac}
\delta \left(S_g+S_{\psi}\right)=0,
\ee
where $S_g=\left(1/16\pi G\right)\int{R\sqrt{-g}d^4x}$ is the standard general relativistic classical action of the gravitational field, while the quantum part of the action is given by
\be\label{9}
S_{\Psi}=\int{\left[{\rm Im}\left \langle \dot{\Psi}|\Psi\right \rangle-\left \langle \Psi |\hat{H}|\Psi \right \rangle +\alpha \left(\left \langle \Psi |\Psi \right \rangle -1\right) \right]dt}.
\ee
  In Eq.~(\ref{9}) $\hat{H}$ is the Hamiltonian operator of the gravitating system, and $\alpha $ is a Lagrange multiplier. The variation  of Eq.~(\ref{quantac}) with respect to the wave function leads to the normalization condition for the quantum wave function $\left<\Psi|\Psi\right>=1$, to the Sch\"odinger equation for $\Psi $,
\be\label{Sch}
i\left|\dot{\Psi}(t)\right>=\hat{H}(t)\left|\Psi (t)\right>-\alpha (t)\left|\Psi (t)\right>,
\ee
and to the semiclassical Einstein equations ~(\ref{gr1}), respectively. Hence, in this simple phenomenological approach to semiclassical gravity, the Bianchi identities still require the conservation of the effective energy-momentum tensor, $\nabla _{\mu}\left<\Psi \right|\hat{T}^{\mu \nu}\left|\Psi \right>=0$.

A more general set of semiclassical Einstein equations, having many similarities with the modified gravity models,  can  be derived by introducing a new coupling between the quantum fields and the classical curvature scalar of the space-time. One such model was proposed in \cite{Kibble}, where the total action containing the geometry-quantum matter coupling term was assumed to be of the form
\be
\int{RF\left(\left<f(\phi)\right>\right)_{\Psi}\sqrt{-g}d^4x},
\ee
where $F$ and $f$ are arbitrary functions, and $\left(\left<f(\phi)\right>\right)_{\Psi}=\left<\Psi (t)\right|f[\phi (x)]\left|\Psi (t)\right>$. Such a geometry-matter coupling term modifies the Hamiltonian $H(t)$ in the Schr\"odinger equation ~(\ref{Sch})  into \cite{Kibble}
\be
\hat{H}(t)\rightarrow \hat{H}_{\Psi}=\hat{H}(t)-\int{N F'\left(\left<f(\phi)\right>\right)_{\Psi}f(\phi)\sqrt{\varsigma}d^3\xi},
\ee
where $N$ is the lapse function, while $\xi ^i$ are intrinsic space-time coordinates, chosen in such a way that the normal vector to a space-like surface is  time-like on the entire space-time manifold.  The scalar function $\varsigma $ can be obtained as  $\varsigma = {\rm det} \;\varsigma _{rs}$, where $\varsigma _{rs}$ is the metric induced on a space-like surface $\sigma( t )$. The surface $\sigma (t)$ globally slices the space-time manifold into
space-like surfaces. Then, by taking into account the effect of the geometry-quantum matter coupling, the effective semiclassical Einstein equations become \cite{Kibble}
\bea\label{123}
R_{\mu \nu}-\frac{1}{2}Rg_{\mu \nu}&=&16\pi G\Big[\left< \hat{T}_{\mu \nu}\right> _{\Psi}+G_{\mu \nu}F-\nonumber\\
&& \nabla _{\mu}\nabla _{\nu} F+
g_{\mu \nu}\Box F\Big].
\eea

In the semi-classical gravitational model introduced  through Eq.~(\ref{123}), the matter energy-momentum tensor is not conserved anymore, since $\nabla _{\mu}\left< \hat{T}^{\mu \nu}\right> _{\Psi}\neq 0$. Thus, describes an effective particle generation process, in which there is a quantum-mechanically induced energy transfer from space-time to matter. Eq.~(\ref{123}). Eq.~(\ref{123}) also gives an effective semiclassical description of the quantum processes in a gravitational field, which are intrinsically related to the matter and energy non-conservation.

It is interesting to compare Eqs.~(\ref{genfe}) and (\ref{123}), both based on some assumptions on the quantum nature of gravity. While Eq.~(\ref{genfe}) is derived through a fist order approximation to quantum gravity, Eq.~(\ref{123}) postulates the existence of a quantum coupling between geometry and matter. While the coupling in Eq.~(\ref{123}) is introduced via a scalar function, the quantum effects are introduced in Eq.~(\ref{genfe}) through the fluctuations of the metric, having a tensor algebraic structure. However, in both models, the matter energy-momentum tensor is generally not conserved, indicating the possibility of the energy transfer between geometry and matter.  Therefore, the physical origin of the modified gravity theories with geometry-matter coupling, which "automatically" leads to  matter creation processes, may be traced back to the semiclassical approximation of the quantum field theory in a Riemannian curved space-time  geometry.

\subsection{The cosmological model}

For cosmological applications we adopt the Friedmann-Robertson-Walker metric,
\be
ds^2=c^2dt^2-a^2(t)\left(\frac{1}{1-Kr^2}dr^2+r^2d\theta^2+r^2sin^2\theta d\phi^2\right),
\ee
where $a(t)$ is the scale factor. The components of the Ricci tensor for this metric are
\begin{eqnarray}
R_{00}=-\frac{3\ddot{a}}{a},
R_{ij}=-\frac{2K+2\dot{a}^2+a\ddot{a}}{a^2}g_{ij}, i,j=1,2,3.
\end{eqnarray}

We define the Hubble function as $H=\dot{a}/a$. As an indicator of the possible accelerated expansion we consider the deceleration
parameter $q$, defined as
\begin{equation}
q=\frac{d}{dt}\frac{1}{H}-1.
\label{deccparam}
\end{equation}
Negative values of $q$ indicate accelerating evolution, while positive ones correspond to decelerating expansion. In
order to perform the comparison between the observational and
theoretical results, instead of the time variable $t$ we introduce the redshift $z$,
defined as
\begin{equation}
1+z=\frac{1}{a},
\label{redshoftfedin}
\end{equation}%
where we have adopted for the scale factor $a\left( z\right) $ the normalization
 $a(t_0)=1 $, where $t_0$ is the present age of the Universe. Hence as a function of the redshift the time derivative operator  can be expressed as
\begin{equation}
\frac{d}{dt}= -H(z)(1+z)\frac{d}{dz}.
\label{timeredshiftrel}
\end{equation}

\section{Modified gravity from quantum perturbation of the metric proportional to the classical metric, \(K^{\mu\nu}=\alpha(x)g^{\mu\nu}\)}\label{sect3}

As a first example of cosmological evolution in modified gravity models induced by the quantum fluctuations of the metric we consider the simple case in which the expectation value of the quantum fluctuation tensor is proportional to the classical metric \cite{re11,re8},
\be\label{K1}
K^{\mu\nu}=\alpha(x)g^{\mu\nu},
\ee
where \(\alpha (x)\) is an arbitrary function of the space-time coordinates $x=x^{\mu}=\left(x^0,x^1,x^2,x^3\right)$. The case $\alpha ={\rm constant}$ was investigated in \cite{re11} and \cite{re8}, respectively. This condition suggest that we have an additional part of the metric, due to the quantum perturbation effects, which is proportional to the classical one.

In the following we will consider two distinct cases, by assuming first than $\alpha (x)$ is a {\it scalar field}, with a specific self-interaction potential. As a second model we will assume that $\alpha (x)$ is a simple scalar function.

\subsection{Scalar field-metric coupling}

If \(\alpha (x)\) is a scalar field, we add an additional Lagrangian
\bea
 L_{\alpha}=\sqrt{-g}\left[\frac{1}{2}\nabla _{\mu }\alpha \nabla ^{\mu}\alpha -V(\alpha)\right],
\eea
as the matter source into the general quantum  perturbed Lagrangian (\ref{2}). Hence we obtain the Lagrangian of our model as
\bea\label{Lalpha}
 L_{total}=&&-\frac{1}{2\kappa^2}\sqrt{-g}\left[(1-\alpha)R\right]+\sqrt{-g}\left(\mathcal{L}_m+\frac{1}{2}\alpha T\right)+\nonumber\\
 &&\sqrt{-g}\left[\frac{1}{2}\nabla _{\mu }\alpha \nabla ^{\mu}\alpha -V(\alpha)\right].
\eea

By taking the variation of the Lagrangian with respect to the field $\alpha $, the Euler-Lagrange equation gives the generalized Klein-Gordon equation for the scalar field, which has the form
\bea\label{KGa}
\Box\alpha-\frac{1}{2\kappa^2}R-\frac{1}{2}T+\frac{\partial V}{\partial \alpha}=0.
\eea
By taking the variation of Eq.~(\ref{Lalpha}) with respect to the metric tensor we obtain the Einstein gravitational field equations as
\bea\label{fealpha}
&&R_{\mu\nu}-\frac{1}{2}R g_{\mu\nu}=\frac{2\kappa ^2}{1-\alpha}\Bigg\{\frac{1+\alpha}{2}T_{\mu\nu}
-\frac{1}{4}\alpha T g_{\mu\nu}+\nonumber\\
&&
\frac{1}{2}\alpha  \theta_{\mu\nu}
-\frac{1}{4}g_{\mu\nu}(\nabla _{\beta }\alpha \nabla ^{\beta}\alpha -2V)+
\frac{1}{2}\nabla_\mu \alpha \nabla_\nu \alpha\Bigg\},\nonumber\\
\eea
where
\bea
\theta_{\mu\nu}= g^{\alpha\beta}\frac{\delta T_{\alpha\beta}}{\delta g^{\mu\nu}}
= -g_{\mu\nu}\mathcal{L}_m-2T_{\mu\nu}.
\eea
After contraction of the Einstein field equations we obtain
\begin{eqnarray}\label{Ralpha}
R=-\kappa ^2\left[T+\frac{\alpha\theta}{1-\alpha}-\frac{1}{1-\alpha}\left(\nabla _{\mu }\alpha \nabla ^{\mu}\alpha -4V\right)\right],
\end{eqnarray}
and thus we can reformulate the gravitational field equations as
\bea
R_{\mu\nu}=&&\frac{\kappa^2}{2(1-\alpha)}\Bigg\{2(1+\alpha)T_{\mu\nu}-Tg_{\mu\nu}+2\alpha\theta_{\mu\nu}\nonumber\\
&&-\alpha\theta g_{\mu\nu}-2Vg_{\mu\nu}+2\nabla_\mu\alpha\nabla_\nu\alpha\Bigg\}.
\eea

The divergence of the matter energy-momentum tensor can be obtained from Eq.~(\ref{fealpha}), and is given by
\bea\label{ex1}
\nabla^\nu T_{\mu\nu}&=&-\frac{1}{2(1+\alpha)} \Bigg\{\alpha\Bigg[2\nabla^\nu\theta_{\mu\nu}-g_{\mu\nu}\nabla^\nu T\Bigg]+ \nonumber\\
&&\frac{\nabla^\nu \alpha}{1-\alpha}\Bigg[4T_{\mu\nu}-Tg_{\mu\nu}+2\theta_{\mu\nu}+\nonumber\\
&&g_{\mu\nu}\left(2V+2(1-\alpha)\Box\alpha-\nabla_\beta \alpha \nabla^\beta \alpha \right)+\nonumber\\
&&2\nabla_\mu \alpha \nabla_\nu \alpha\Bigg]+2\nabla^\nu Vg_{\mu\nu}\Bigg\}.
\eea
In the following we will adopt for the matter energy-momentum tensor the perfect fluid form
\begin{eqnarray}\label{28}
T_{\mu\nu}=\left(\rho +p\right)u_\mu u_\nu -g_{\mu\nu}p,
\end{eqnarray}
where $\rho$ is the matter energy density, $p$ is the thermodynamic pressure, and $u^{\mu}$ is the matter four velocity, satisfying the normalization condition $u_{\mu}u^{\mu}=1$. For this choice of the energy-momentum tensor we have

\bea
\theta_{\mu\nu}&=& g^{\alpha\beta}\frac{\delta T_{\alpha\beta}}{\delta g^{\mu\nu}}
= -g_{\mu\nu}\mathcal{L}_m-2T_{\mu\nu}=\nonumber\\
&&g_{\mu\nu}p-2\left(\rho +p\right)u_\mu u_\nu .
\eea

To obtain the above equation we have adopted for the matter Lagrangian the representation \(\mathcal{L}_m=p\). For the scalar $\theta $ we obtain
\be
\theta=2\left(p-\rho \right).
\ee

With the use of Eq.~(\ref{Ralpha}) the Klein-Gordon equation now reads
\bea
\Box\alpha+\frac{1}{2(1-\alpha)}\left(\alpha\theta-\nabla _{\mu }\alpha \nabla ^{\mu}\alpha +4V\right)+\frac{\partial V}{\partial \alpha}=0.
\eea

\subsubsection{Cosmological applications}

In the following we will restrict our analysis to the case of the flat Friedmann-Robertson-Walker metrics, and hence we set \(K=0\) in the gravitational field equations. The Friedmann and the Klein-Gordon equations describing our generalized gravity model obtained from a fluctuating metric take the form
\be\label{f1alpha}
3H^2=\frac{\kappa ^2}{1-\alpha}\Bigg\{\rho-\frac{1}{2}\alpha(3\rho-p)+\frac{1}{2}\dot{\alpha}^2+V\Bigg\},
\ee
\be\label{f2alpha}
2\dot{H}+3H^2=\frac{-\kappa ^2}{1-\alpha}\Bigg\{p+\frac{\alpha}{2}(\rho -3p)+\frac{1}{2}\dot{\alpha}^2-V\Bigg\},
\ee
\bea\label{f3alpha}
\ddot{\alpha}+3H\dot{\alpha} +\frac{1}{2(1-\alpha)}\left[2\alpha \left(p-\rho \right)-\dot{\alpha}^2+4V\right]\frac{\partial V}{\partial \alpha}=0,\nonumber\\
\eea
where
\bea\label{35}
\hspace{-0.05cm}\Box\alpha=\nabla_\nu\nabla^\nu\alpha&=&\frac{1}{\sqrt{-g}}\frac{\partial(\sqrt{-g}\nabla^\nu\alpha)}{\partial x^\nu}
=\Bigg(\ddot{\alpha}+\dot{\alpha}\frac{3\dot{a}}{a}\Bigg).\nonumber\\
\eea

\paragraph{The energy conservation equation}

By multiplying Eq.~(\ref{f1alpha}) with \(a^3\),  and taking the time derivative of  its both sides, we obtain
\bea
3H^2+6\frac{\ddot{a}}{a}=\frac{\kappa^2}{1-\alpha}\Bigg\{3\Bigg[\rho-\frac{\alpha}{2}(3\rho-p)+\frac{\dot{\alpha}^2}{2}+V\Bigg]\nonumber\\
+\frac{a}{\dot{a}}\Bigg[\dot{\rho}-\frac{\dot{\alpha}}{2}(3\rho -p)-\frac{\alpha}{2}(3\dot{\rho}-\dot{p})
+\dot{\alpha}\ddot{\alpha}+\frac{\partial V}{\partial \alpha}\dot{\alpha}\Bigg]\nonumber\\
+\frac{a}{\dot{a}}\frac{\dot{\alpha}}{1-\alpha}\Bigg[\rho -\frac{1}{2}\alpha(3\rho -p)+\frac{1}{2}\dot{\alpha}^2+V\Bigg]\Bigg\}.\nonumber\\
\eea
With the use of Eq.~(\ref{f2alpha}) we obtain
\bea\label{encons1}
\Bigg[(1-\alpha)(\rho+p)+\dot{\alpha}^2\Bigg]\frac{d}{dt}a^3+\left(1-\frac{3}{2}\alpha\right)a^3\frac{d}{dt}\rho\nonumber\\
=-\frac{a^3}{2}\Bigg\{\alpha\dot{p}-\dot{\alpha}(3\rho-p)+2\dot{\alpha}\ddot{\alpha}+2\frac{\partial V}{\partial \alpha}\dot{\alpha}\nonumber\\
+\frac{2\dot{\alpha}}{1-\alpha}\Bigg[\rho-\frac{1}{2}\alpha(3\rho-p)+\frac{1}{2}\dot{\alpha}^2+V\Bigg]\Bigg\}.\nonumber\\
\eea

Eq.~(\ref{encons1}) can be rewritten in the equivalent form
\bea\label{constan1}
\frac{d}{dt}[(1-\alpha)\rho a^3]+(1-\alpha)p\frac{d}{dt}a^3=\frac{\alpha}{2}a^3(\dot{\rho}-\dot{p})\nonumber\\
-\frac{a^3}{2}\frac{\dot{\alpha}}{1-\alpha}\Bigg[p-\rho+2V+\dot{\alpha}^2\Bigg]-a^3\frac{\partial V}{\partial \alpha}\dot{\alpha}-\nonumber\\
a^3(\dot{\alpha}\ddot{\alpha}+\dot{\alpha}^2\frac{3\dot{a}}{a})-a^3\rho\dot{\alpha}.
\eea

The same conservation equation can be obtained directly from the field equations (\ref{ex1}). By taking into account the explicit form of the matter energy-momentum tensor as given by Eq.~(\ref{28}),  and that
\bea
\theta_{\mu\nu}=g_{\mu\nu}p-2(\rho+p)u_\mu u_\nu,
\eea
we have
\bea
\nabla^\nu T_{\mu\nu}=\dot{\rho}+3(\rho+p)\frac{\dot{a}}{ca},
\eea
and
\bea
\nabla^\nu \theta_{\mu\nu}=-2\dot{\rho}-\dot{p}-6(\rho+p)\frac{\dot{a}}{ca},
\eea
respectively. Hence Eq.~(\ref{28}) takes the form
\bea
-2(1-\alpha)\left(\rho+p\right)\frac{3\dot{a}}{a}-2\left(1-\frac{3}{2}\alpha\right)\dot{\rho}-\alpha\dot{p}=2\frac{\partial V}{\partial \alpha}\dot{\alpha}\nonumber\\
+\frac{\dot{\alpha}}{1-\alpha}\Bigg[p-\rho+2V+2(1-\alpha)\Box\alpha+\dot{\alpha}^2\Bigg].\nonumber\\
\eea

By taking into account Eq.~(\ref{35}), after eliminating $\Box\alpha$ from the above equation,  we reobtain again the conservation equation Eq.~(\ref{constan1}).

\paragraph{The dimensionless form of the cosmological evolution equation}

In order to simplify the mathematical formulation of the cosmological model  we rescale first the field $\alpha $ and its potential $V$ as $\alpha \rightarrow \kappa \alpha$, and $V\rightarrow \kappa ^2V$, respectively. Next, we introduce a set of dimensionless variables $\left(\tau, h,r,P,v\right)$, defined as
\be
\tau =H_0t,H=H_0h,\rho =\frac{3H_0^2}{8\pi G}, p=\frac{3H_0^2}{8\pi G}P,v=\frac{1}{H_0^2}V,
\ee
where $H_0$ is the present day value of the Hubble function.  Then the cosmological evolution equation take the dimensionless form
\be\label{d1}
3h^2=\frac{1}{1-\alpha}\left[r-\frac{1}{2}\alpha (3r-P)+\frac{1}{2}\left(\frac{d\alpha }{d\tau}\right)^2+v\right],
\ee
\bea\label{d2}
2\frac{dh}{d\tau}+3h^2&=&\frac{1}{1-\alpha }\Bigg\{-\left[P+\frac{\alpha}{2}\left(r-3P\right)\right]-\nonumber\\
&&\left[\frac{1}{2}\left(\frac{d\alpha }{d\tau}\right)^2-v\right]\Bigg\},
\eea
\bea\label{d3}
&&\frac{d^2\alpha }{d\tau ^2}+3h\frac{d\alpha }{d\tau}+\frac{1}{2(1-\alpha)}\Bigg[2\alpha (P-r)-\left(\frac{d\alpha }{d\tau}\right)^2+\nonumber\\
&&4V\Bigg]+
\frac{\partial v}{\partial \alpha}=0.
\eea

In order to close the system of Eqs.~(\ref{d1})-{\ref{d3}) we must specify the equation of state of the matter $P=P(r)$, and the functional form of the self-interaction potential of the scalar field $v$. In the following we will restrict our analysis to the case of the dust, with $P=0$. Then, by denoting $u=d\alpha /d\tau$, from Eq.~(\ref{d1}) we obtain for the dimensionless energy density $r$ the expression
\be
r=\frac{1}{1-3\alpha /2}\left[3(1-\alpha )h^2-\left(\frac{1}{2}u^2+v\right)\right].
\ee
Then, by introducing the redshift $z$ as an independent variable, it follows that the cosmological evolution is described by the following system of equations,
\be\label{daz1}
\frac{d\alpha }{dz}=-\frac{1}{1+z}\frac{u}{h},
\ee
\be\label{dhz1}
\frac{dh}{dz}=\frac{1}{2(1+z)h}\left[\frac{u^2}{1-\alpha }-\frac{6 (\alpha -1) h^2+u^2+2 v}{2\left(1-3 \alpha /2\right)}\right],
\ee
\bea\label{duz1}
\frac{du}{dz}&=&\frac{3}{1+z}u-\frac{4 (\alpha -1) \left(2 v-3 \alpha  h^2\right)+(2-5 \alpha ) u^2}{4(1+z)(1-\alpha)(1-3 \alpha/2)h}+\nonumber\\
&&\frac{1}{(1+z)h}\frac{\partial v}{\partial \alpha}.
\eea

For the deceleration parameter we obtain the expression
\bea
q&=&(1+z)\frac{1}{h}\frac{dh}{dz}-1=\nonumber\\
&&\frac{1}{h^2}\left[\frac{u^2}{1-\alpha }-\frac{6 (\alpha -1) h^2+u^2+2 v}{2\left(1-3 \alpha /2\right)}\right]-1.
\eea

For the scalar field self-interaction potential we adopt a Higgs type form, so that
\be\label{potH1}
v(\alpha)=\frac{\mu ^2}{2}\alpha ^2-\frac{\lambda }{4}\alpha ^4,
\ee
where $\mu ^2>0$ and $\lambda >0$ are constants.
 In the following we will consider two cases. In the first case we assume that the Universe dominated by the quantum fluctuations of the metric evolves in the minimum of the Higgs potential, so that $\partial v/\partial \alpha =0$, implying $\alpha =\pm \sqrt{\mu ^2/\lambda}$, and $v(\alpha)=\mu ^4/4\lambda ={\rm constant}$. Secondly, we will investigate the evolution of the Universe in the presence of a time varying "full" Higgs type scalar field self-interaction potential (\ref{potH1}). Once the form of the potential is fixed, the system of differential equations Eqs.~(\ref{daz1})-(\ref{duz1}) must be integrated with the initial conditions $\alpha (0)=\alpha _0$, $h(0)=1$, and $u(0)=u_0$, respectively.

 \paragraph{Cosmological evolution of the Universe in the minimum of the Higgs potential}

In Figs.~\ref{fig1}-\ref{fig4} we present the results of the numerical integration of the system of cosmological evolution equations Eqs.~(\ref{daz1})-(\ref{duz1}), for different values of the constant self-interaction potential $v=v_0={\rm constant}$. The initial values of $\alpha $ and $u$ used to integrate the system are $\alpha (0)=0.01$ and $u(0)=0.1$, respectively.

\begin{figure}[tbp]
\includegraphics[width=7.5cm, angle=0]{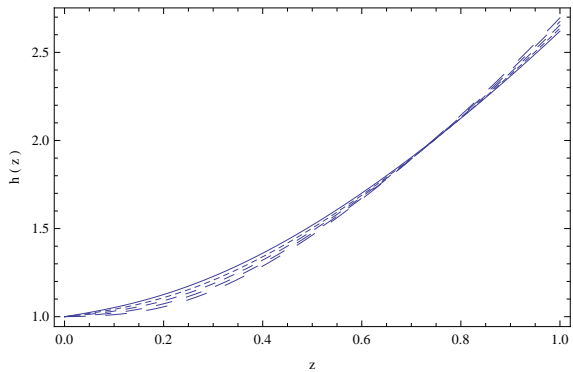} %
\caption{Variation with respect to the redshift of the dimensionless Hubble function for  the Universe in the modified gravity model induced by the coupling between the metric and a Higgs type scalar field, in the presence of a constant self-interaction potential $v=v_0$, for different values of $v_0$: $v_0=2.1$ (solid curve), $v_0=2.3$ (dotted curve), $v_0=2.5$ (short dashed curve), $v_0=2.7$ (dashed curve), and $v_0=2.9$ (long dashed curve), respectively.}
\label{fig1}
\end{figure}

\begin{figure}[tbp]
\includegraphics[width=7.5cm, angle=0]{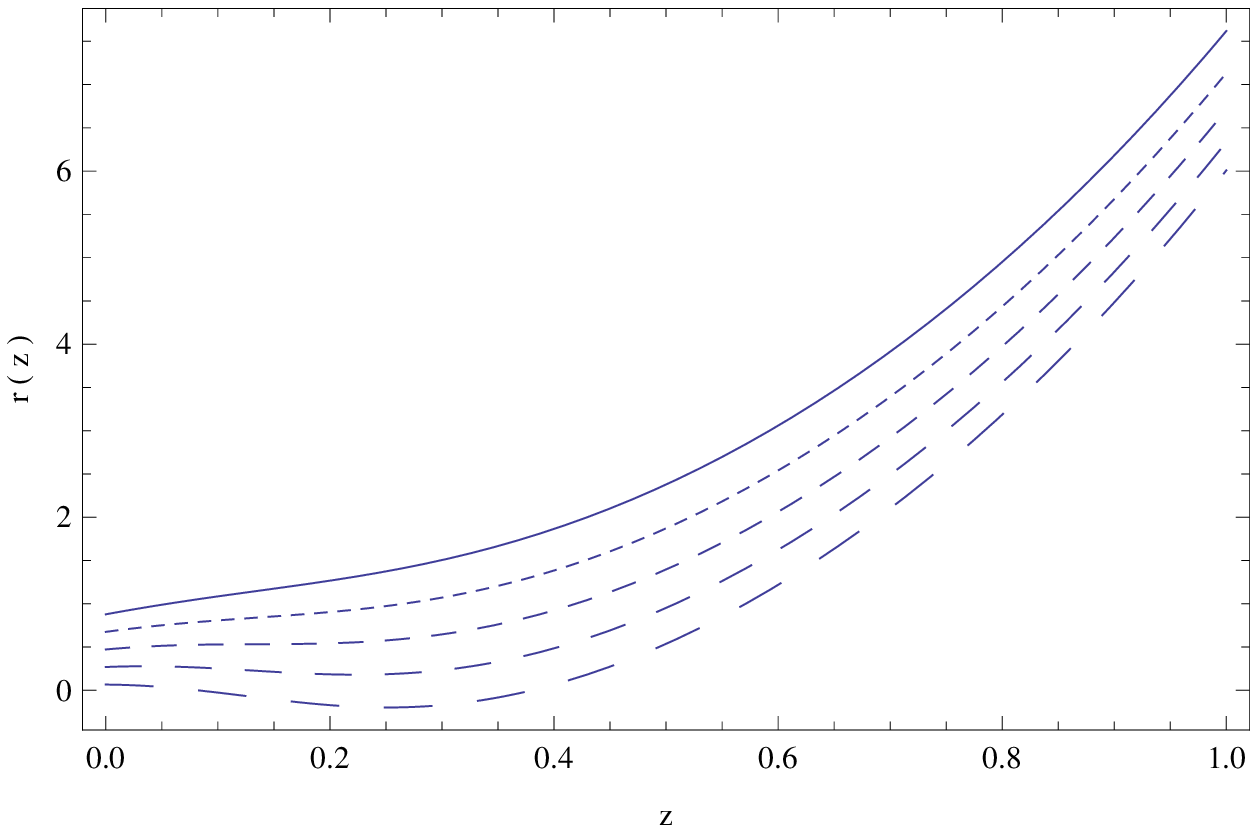} %
\caption{Variation with respect to the redshift of the dimensionless matter energy density for  the Universe in the modified gravity model induced by the coupling between the metric and a Higgs type scalar field, in the presence of a constant self-interaction potential $v=v_0$, for different values of $v_0$: $v_0=2.1$ (solid curve), $v_0=2.3$ (dotted curve), $v_0=2.5$ (short dashed curve), $v_0=2.7$ (dashed curve), and $v_0=2.9$ (long dashed curve), respectively. }
\label{fig2}
\end{figure}

\begin{figure}[tbp]
\includegraphics[width=7.5cm, angle=0]{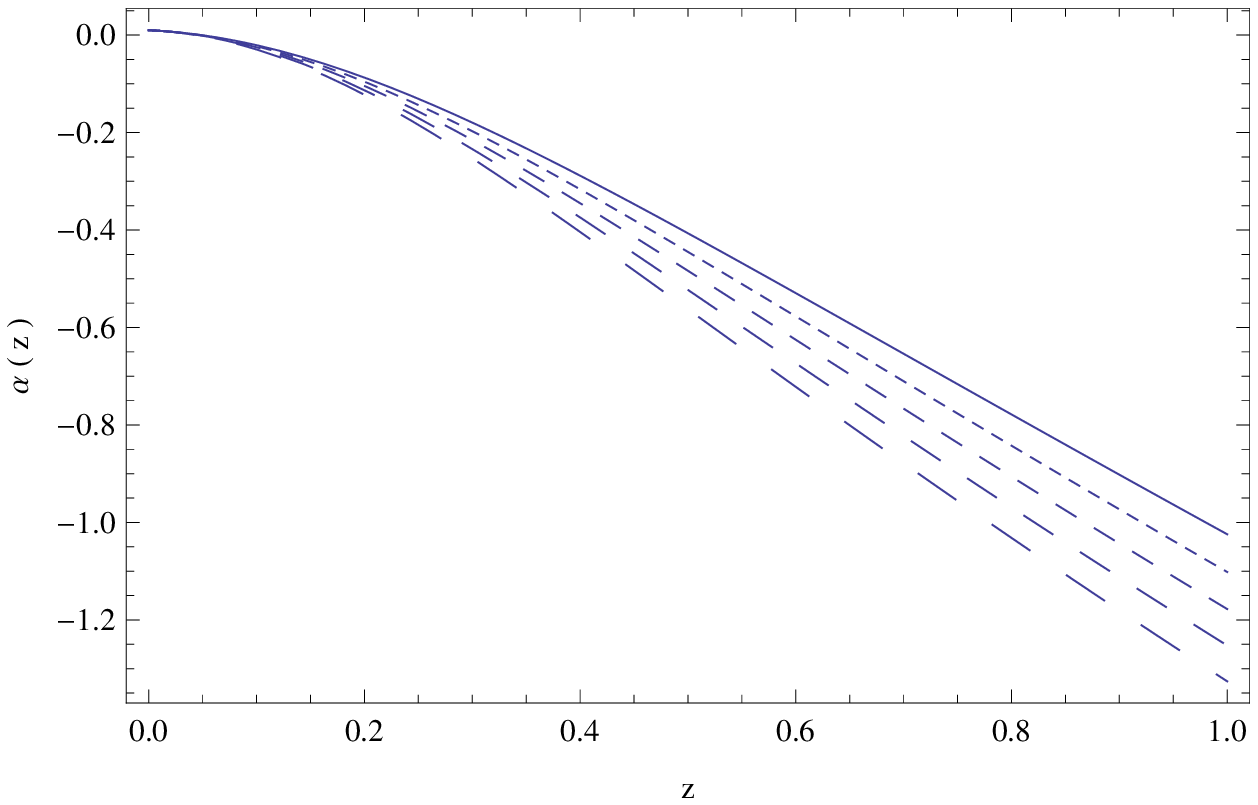} %
\caption{Variation with respect to the redshift of the scalar field $\alpha$  for  the Universe in the modified gravity model induced by the coupling between the metric and a Higgs type scalar field, in the presence of a constant self-interaction potential $v=v_0$, for different values of $v_0$: $v_0=2.1$ (solid curve), $v_0=2.3$ (dotted curve), $v_0=2.5$ (short dashed curve), $v_0=2.7$ (dashed curve), and $v_0=2.9$ (long dashed curve), respectively.}
\label{fig3}
\end{figure}

\begin{figure}[tbp]
\includegraphics[width=7.5cm, angle=0]{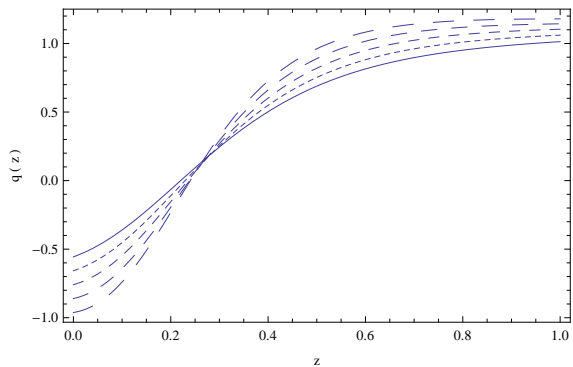} %
\caption{Variation with respect to the redshift of the deceleration parameter $q$ for  the Universe in the modified gravity model induced by the coupling between the metric and a Higgs type scalar field, in the presence of a constant self-interaction potential $v=v_0$, for different values of $v_0$: $v_0=2.1$ (solid curve), $v_0=2.3$ (dotted curve), $v_0=2.5$ (short dashed curve), $v_0=2.7$ (dashed curve), and $v_0=2.9$ (long dashed curve), respectively.   }
\label{fig4}
\end{figure}

The Hubble function, represented in Fig.~\ref{fig1}, is a monotonically increasing function of the redshift (a monotonically decreasing function of the cosmological time), indicating an expansionary evolution of the Universe. Its variation is basically independent on the adopted numerical values of $v_0$, and, at $z\in [0,0.10]$, $h$ becomes approximately constant, indiating that the Universe has entered in a de Sitter type phase. The matter energy density, depicted in Fig.~\ref{fig2}, is a monotonically increasing function of the redshift, and its variation show a strong dependence on the numerical value of $v_0$. In the considered range of values of $v_0$ at the present time the matter density can either reach values of the order of the critical density, with $r(0)\approx 1$, or become negligibly small. The function $\alpha $, shown in Fig.~\ref{fig3}, monotonically decreases with the redshift, and has negative numerical values. For large redshifts, the variation of $\alpha $ has a strong dependence on the numerical values of $v_0$, However, for $z\leq 0.2$, the changes in $\alpha $ due to the variation of $v_0$ become negligible, and for $z\in [0,0.05]$ $\alpha $ becomes a constant. The deceleration parameter $q$, plotted in Fig.~\ref{fig4}, shows a complex behavior. The Universe starts its evolution at a redshift $z=1$ in a decelerating phase, with $q\approx 1$. The Universe begins to accelerates, and it enters in a marginally accelerating phase, with $q=0$, at a redshift of around $z\approx 0.3$. The variation of the deceleration parameter strongly depends on the numerical values of $v_0$, and, depending on this numerical value, the evolution of the Universe at the present time can either be de Sitter, with $q=-1$, or with higher values of $q$, of the order of $q\approx -0.5$.

 \paragraph{Cosmological evolution in the presence of the Higgs potential}

 The evolution of the cosmological and physical parameters of a Universe filled with a scalar field with Higgs potential (\ref{potH1}), coupled to the fluctuating quantum metric, are presented in Figs.~\ref{fig5}-\ref{fig8}. To numerically integrate the gravitational field equations (\ref{daz1})-(\ref{duz1}) in the redshift range $z\in [0,2]$ we have adopted the initial conditions $\alpha (0)=0.1$ and $u(0)=0.1$, respectively. We have fixed the value of the coefficient $\lambda $ in the Higgs potential as $\lambda =150$, and we have varied the mass $\mu ^2>0$ of the Higgs field.

 \begin{figure}[tbp]
\includegraphics[width=7.5cm, angle=0]{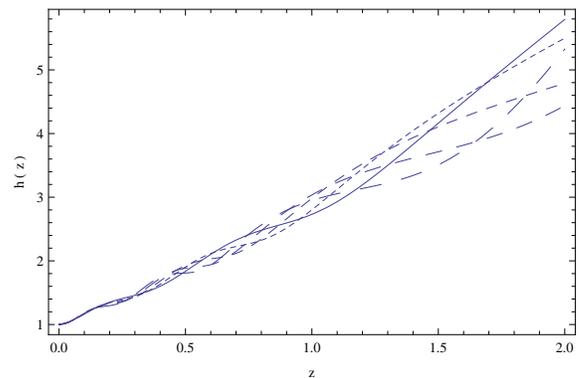} %
\caption{Variation with respect to the redshift of the dimensionless Hubble function for  the Universe filled with a Higgs type scalar field coupled to the fluctuating quantum metric for  $\lambda =150$, and for different values of $\mu ^2$: $\mu ^2=250$ (solid curve), $\mu ^2=300$ (dotted curve), $\mu ^2=350$ (short dashed curve), $\mu ^2=400$ (dashed curve), and $\mu ^2=450$ (long dashed curve), respectively.}
\label{fig5}
\end{figure}

\begin{figure}[tbp]
\includegraphics[width=7.5cm, angle=0]{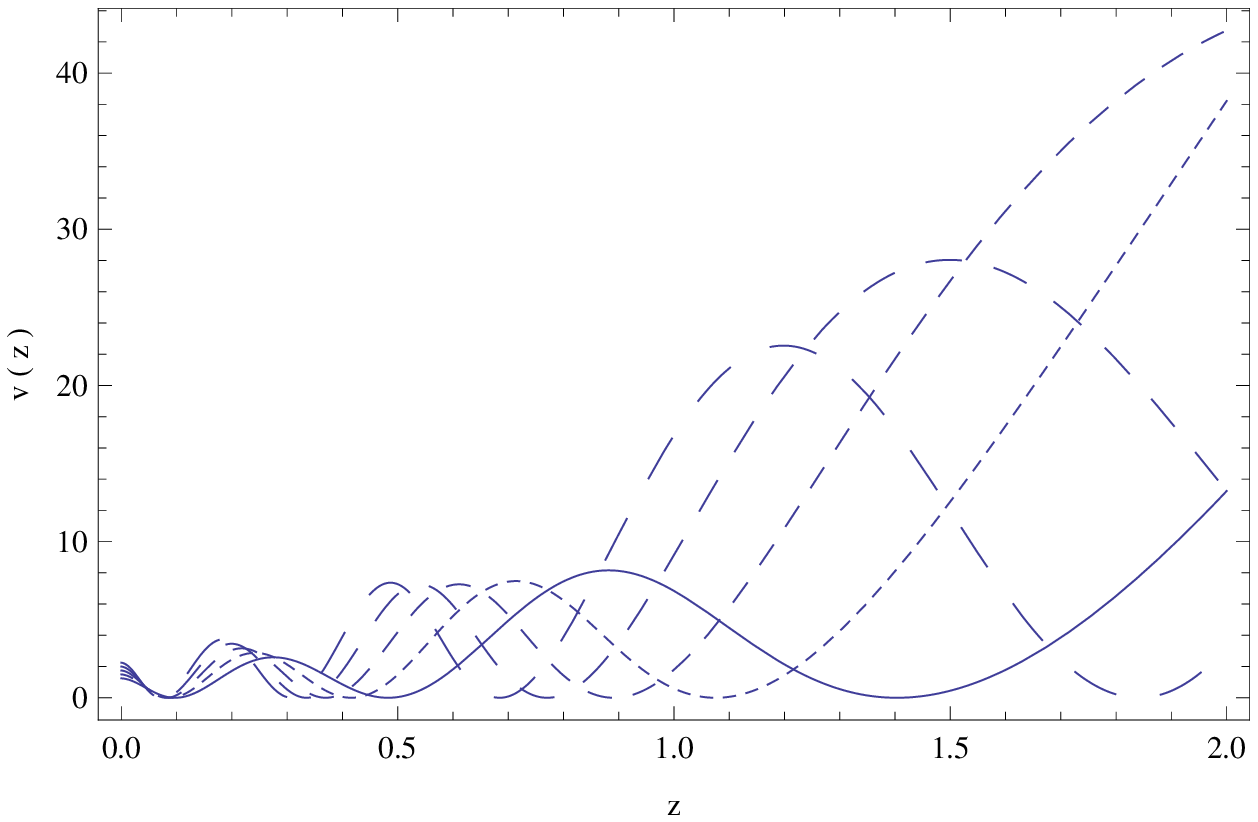} %
\caption{Variation with respect to the redshift of the potential $v$ in the modified gravity model induced by the coupling between the metric and a Higgs type scalar field for  $\lambda =150$, and for different values of $\mu ^2$: $\mu ^2=250$ (solid curve), $\mu ^2=300$ (dotted curve), $\mu ^2=350$ (short dashed curve), $\mu ^2=400$ (dashed curve), and $\mu ^2=450$ (long dashed curve), respectively. }
\label{fig6}
\end{figure}

\begin{figure}[tbp]
\includegraphics[width=7.5cm, angle=0]{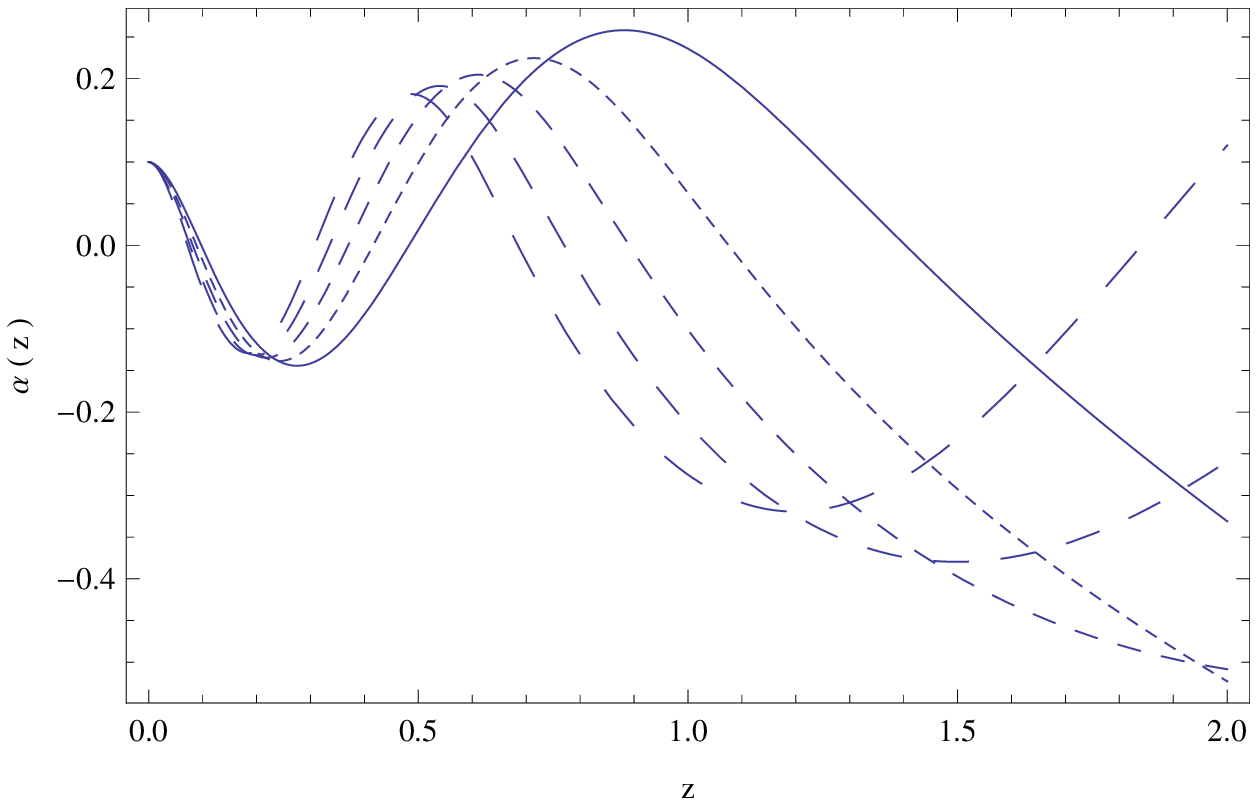} %
\caption{Variation with respect to the redshift of the scalar field $\alpha$ in the modified gravity model induced by the coupling between the metric and a Higgs type scalar field, for  $\lambda =150$, and for different values of $\mu ^2$: $\mu ^2=250$ (solid curve), $\mu ^2=300$ (dotted curve), $\mu ^2=350$ (short dashed curve), $\mu ^2=400$ (dashed curve), and $\mu ^2=450$ (long dashed curve), respectively.}
\label{fig7}
\end{figure}

\begin{figure}[tbp]
\includegraphics[width=7.5cm, angle=0]{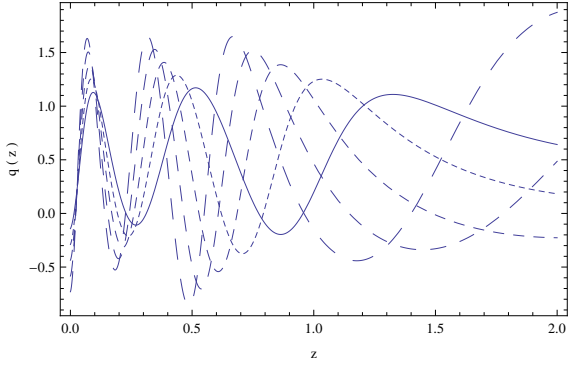} %
\caption{Variation with respect to the redshift of the deceleration parameter $q$ in the modified gravity model induced by the coupling between the metric and a Higgs type scalar field, for a $\lambda =150$, and for different values of $\mu ^2$: $\mu ^2=250$ (solid curve), $\mu ^2=300$ (dotted curve), $\mu ^2=350$ (short dashed curve), $\mu ^2=400$ (dashed curve), and $\mu ^2=450$ (long dashed curve), respectively.}
\label{fig8}
\end{figure}

The Hubble function, plotted in Fig.~\ref{fig5}, is an increasing function of the redshift, indicating an expansionary evolution. However, $h$ has a complex behavior, which is strongly dependent on the numerical value of $\mu ^2$. The variation with $z$ of the Higgs potential is represented in Fig.~\ref{fig6}. The potential has a damped harmonic oscillator type behavior, with the amplitude of the oscillations decreasing while we are approaching the present day moment, with $z=0$. The same damped oscillator type pattern can be observed in the redshift evolution of the scalar field $\alpha $, depicted in Fig.~\ref{fig7}. An oscillatory behavior is also characteristic for the variation with respect to $z$  of the deceleration parameter $q$, presented in Fig.~\ref{fig8}. The behavior is strongly dependent on the numerical values of $\mu ^2$, so that at $z=2$ the Universe can be in either a decelerating ($q.0$), or in an accelerating phase, with $q<0$. There is an alternation of the accelerating and decelerating epochs, but, (almost) independently of the values of $\mu ^2$, the Universe enters a very rapid accelerating phase at $z\approx 0.10$, which brings down in a very short cosmological time interval the deceleration parameter from $q\approx 1.5$ to $q\approx -1$. Hence the de Sitter solution is also an attractor of this model.

\subsection{Scalar function coupling to the metric}

As a second example of modified gravity model induced by a quantum perturbation tensor of the form (\ref{K1}) we assume that $alpha $ is a "pure" function of the coordinates, and there is no specific physical field associated to it. Then,
by using the least action principle, from Eq.~(\ref{2}) we obtain the gravitational field equations as
\bea\label{4}
R_{\mu\nu}-\frac{1}{2}R g_{\mu\nu}&=&\frac{2\kappa ^2}{1-\alpha(x)}\Bigg\{\frac{1}{2}\left[1+\alpha(x)\right]T_{\mu\nu}-\nonumber\\
&&\frac{1}{4}\alpha (x)T g_{\mu\nu}+\frac{1}{2}\alpha (x) \theta_{\mu\nu}\Bigg\},
\eea
 where we have denoted $\theta_{\mu\nu}= g^{\alpha\beta}\left(\delta T_{\alpha\beta}/\delta g^{\mu\nu}\right)$, and
  \be
  \frac{\delta T}{\delta g^{\mu\nu}}=T_{\mu\nu}+\theta_{\mu\nu},
  \ee
 respectively. As usual, \(T=T_{\mu\nu}g^{\mu\nu}\) is the trace of the energy-momentum tensor, and we have denoted \(\theta =\theta_{\mu\nu}g^{\mu\nu}\). After contraction Eq~(\ref{4}) reads
\begin{eqnarray}\label{7}
R=-\kappa ^2\left[T+\frac{\alpha(x)}{1-\alpha (x)}\theta\right].
\end{eqnarray}
Combining Eqs.~(\ref{4}) and (\ref{7}) we obtain the field equations in the form
\begin{eqnarray}
R_{\mu\nu}&=&\frac{2\kappa ^2}{1-\alpha (x)}\Bigg[\frac{1}{2}(1+\alpha (x))T_{\mu\nu}-\frac{1}{4} T g_{\mu\nu}+ \nonumber\\
&&\frac{1}{2}\alpha (x) \theta_{\mu\nu}-\frac{1}{4}\alpha (x)\theta g_{\mu\nu}\Bigg].
\end{eqnarray}

Assuming that the Lagrangian density of the matter \(\mathcal{L}_m\) depends only on the metric tensor but not on its derivatives, we obtain
\begin{eqnarray}
T_{\mu\nu}=2\frac{\partial \mathcal{L}_m}{\partial g^{\mu\nu}}-g_{\mu\nu}\mathcal{L}_m,
\end{eqnarray}
and
\begin{eqnarray}
\theta_{\mu\nu}=2g^{\alpha\beta}\frac{\partial^2 \mathcal{L}_m}{\partial g^{\alpha\beta}\partial g^{\mu\nu}}-g_{\mu\nu}\mathcal{L}_m-2T_{\mu\nu},
\end{eqnarray}
respectively. Different kinds of matter Lagrangians may lead to different \(\theta_{\mu\nu}\), and therefore to different gravitational theories.

By taking into account the mathematical identity \(\nabla^\nu G_{\mu\nu}=0\), after taking the covariant divergence of Eq.~(\ref{4}) we obtain
\begin{eqnarray}
&&0=\frac{\kappa ^2\left(\nabla^\nu \alpha\right)}{2(1-\alpha)^2} \left[2(1+\alpha)T_{\mu\nu}-\alpha T g_{\mu\nu}+2\alpha \theta_{\mu\nu}\right]+ \nonumber\\
&&\frac{\kappa ^2}{2(1-\alpha)}\Bigg[ 2\left(\nabla^\nu \alpha\right) T_{\mu\nu}+2(1+\alpha)\nabla^\nu T_{\mu\nu}-\left(\nabla^\nu\alpha\right) \times \nonumber\\
&&T g_{\mu\nu}-\alpha (\nabla^\nu T) g_{\mu\nu}+2\left(\nabla^\nu\alpha\right) \theta_{\mu\nu}+2\alpha (\nabla^\nu\theta_{\mu\nu})\Bigg]=\nonumber\\
&&\left(\nabla^\nu \alpha\right)\Bigg[\frac{2(1+\alpha)T_{\mu\nu}-\alpha T g_{\mu\nu}+2\alpha \theta_{\mu\nu}}{1-\alpha}+2T_{\mu\nu}-\nonumber\\
&&Tg_{\mu\nu}+2\theta_{\mu\nu}\Bigg]+\alpha\Bigg[2\nabla^\nu T_{\mu\nu}-g_{\mu\nu}\nabla^\nu T+2\nabla^\nu\theta_{\mu\nu}\Bigg]+\nonumber\\
&&2\nabla^\nu T_{\mu\nu}.
\end{eqnarray}

Now it is easy to check that the divergence of the matter energy-momentum tensor is
\begin{eqnarray}\label{emtc}
\nabla^\nu T_{\mu\nu}&=&-\frac{1}{2(1+\alpha)} \Bigg\{\left(\nabla^\nu \alpha\right)\Bigg[\frac{4T_{\mu\nu}-T g_{\mu\nu}+2\theta_{\mu\nu}}{1-\alpha}\Bigg]+\nonumber\\
&&\alpha\Bigg[-g_{\mu\nu}\nabla^\nu T+2\nabla^\nu\theta_{\mu\nu}\Bigg]\Bigg\}.
\end{eqnarray}

In the following discussion, we consider a perfect fluid described by
$\rho $, the matter energy density, and by $p$, the thermodynamic pressure. In the comoving frame with $u^{\mu}=(1,0,0,0)$ the components of the energy-momentum tensor have the components $T^{\mu}_{\nu}={\rm diag}\left(\rho,-p,-p,-p\right)$.
Moreover, we will consider that matter obeys an equation of state of the form $p=\omega \rho$, where $\omega ={\rm constant}$, and $0\leq \omega \leq 1$. Then the generalized Friedmann equations that follow from the gravitational field equations Eqs.~(\ref{4}) are
\be\label{f1c}
H^2=\frac{\kappa ^2}{3}\left[\frac{2-(3-\omega)\alpha}{2\left(1-\alpha\right)}\right]\rho,
\ee
\be\label{f2c}
\frac{\ddot{a}}{a}=-\frac{\kappa ^2}{6}\left[\frac{1+(3-4\alpha)\omega}{1-\alpha}\right]\rho,
\ee
or, equivalently,
\be\label{dda}
\frac{\ddot{a}}{a}=-\frac{1+(3-4\alpha)\omega}{2-(3-\omega)\alpha}H^2.
\ee
Eq.~(\ref{f2c}) can be alternatively written as
\be\label{f3c}
2\dot{H}+3H^2=-\kappa ^2\left[\frac{2\omega+\alpha(1-3\omega)}{2\left(1-\alpha\right)}\right]\rho.
\ee
Hence
\be\label{f4c}
\dot{H}=-\frac{\kappa ^2}{2}\left(1+\omega \right)\rho =\frac{3 (\alpha -1) (\omega +1)}{\alpha  (\omega -3)+2} H^2.
\ee
For the deceleration parameter we obtain
\be
q=\frac{-\alpha  (7 \omega +3)+6 \omega +4}{\alpha  (\omega -3)+2}.
\ee
 The presence of an accelerated expansion requires $\ddot{a}>$, which imposes on the function $\alpha (t)$ the condition
 \be
\frac{\alpha (t)}{1-\alpha (t)}>\frac{1+3\omega }{4\omega }\frac{1}{1-\alpha (t)}, \forall t\geq t_a.
\ee

\subsubsection{Conservative models-the \(\nabla^\nu T_{\mu\nu}=0\) case}

In standard general relativity theory, the energy-momentum tensor is conserved, (\(\nabla^\nu T_{\mu\nu}=0\)), while in modified theories of gravity the classical-defined energy-momentum tensor is not always conserved. However, in the modified gravity theory induced by the quantum fluctuations of the metric, with expectation value of the fluctuations proportional to the metric tensor,  thanks to the addition of the function $\alpha $, we can maintain the conservation of the energy and momentum by constraining the new function. In the following we  assume for the matter Lagrangian the form  \(L_m=p\), and, since the second derivatives of \(L_m\) with respect to the metric tensor are zero, we obtain for the tensor
$\theta _{\mu \nu}$ the expression
\be
\theta_{\mu\nu}=-g_{\mu\nu}L_m-2T_{\mu\nu}.
\ee
 Demanding that \(\nabla^\nu T_{\mu\nu}=0\) we obtain
\begin{eqnarray}\label{20}
&&0=\left(\nabla^\nu \alpha\right)\Bigg[\frac{2(1+\alpha)T_{\mu\nu}-\alpha T g_{\mu\nu}+2\alpha \theta_{\mu\nu}}{1-\alpha}+2T_{\mu\nu}-\nonumber\\
&&Tg_{\mu\nu}+2\theta_{\mu\nu}\Bigg]
+\alpha\Bigg[2\nabla^\nu\theta_{\mu\nu}-g_{\mu\nu}\nabla^\nu T\Bigg].
\end{eqnarray}

Multiplying by \(u^{\mu}\) both sides of Eq~(\ref{20}) we obtain
\begin{eqnarray}
\dot{\alpha}\left[\frac{-\rho +p}{1-\alpha}\right]+
\alpha\left[-5\dot{\rho}+\dot{p}\right]=0,
\end{eqnarray}
giving
\begin{eqnarray}
\frac{\dot{\alpha}}{\alpha(1-\alpha)}&=&\frac{-5\dot{\rho}+\dot{p}}{\rho-p}.
\end{eqnarray}
For the linear barotropic equation of state with $p=\omega \rho $, for $\omega \neq 1$ we obtain
\be
\frac{\dot{\alpha}}{\alpha(1-\alpha)}=-\left(\frac{5-\omega}{1-\omega}\right)\frac{\dot{\rho}}{\rho},
\ee
which gives the density as a function of $\alpha $ as
\be\label{24}
\rho =\rho_0 \left(\frac{\alpha}{1-\alpha}\right)^{-\frac{1-\omega}{5-\omega}},
\ee
where $\rho _0$ is an arbitrary constant of integration. On the other hand the conservation of the energy-momentum tensor gives the equation
\be
-3\frac{\dot{a}}{a}(1+\omega)=\frac{\dot{\rho}}{\rho},
\ee
which give for the matter density the standard relation
\be\label{26}
\rho=\rho_0^{\prime}a^{-3(1+\omega)},
\ee
where $\rho_0^{\prime}$ is an arbitrary constant of integration. From Eqs.~(\ref{24}) and (\ref{26}) we obtain the scale factor dependence of the function $\alpha $ as
\be
\frac{\alpha }{1-\alpha }=\alpha _0a^{3n},
\ee
where $n=(1+\omega)(5-\omega)/(1-\omega)$, and $\alpha _0=\left(\rho_0^{\prime}/\rho _0\right)^{-(5-\omega)/(1-\omega)}$.
The the first Friedmann equation Eq.~(\ref{f1c}) gives
\be
\frac{\dot{a}}{a}=\frac{\kappa}{\sqrt{6}} \sqrt{\rho _0 a^{-3 (\omega +1)} \left(\alpha _0 (\omega -1)
   a^{3 n}+2\right)},
\ee
from which we obtain
\bea
\frac{\kappa }{\sqrt{6}}\left( t-t_{0}\right) &=&\frac{\sqrt{2\alpha
_{0}(\omega -1)a^{3n}+4}}{3(\omega +1)}\times \nonumber\\
&&\frac{\,_{2}F_{1}\left( \frac{1}{2},%
\frac{\omega +1}{2n};\frac{\omega +1}{2n}+1;-\frac{1}{2}a^{3n}\alpha
_{0}(\omega -1)\right) }{\sqrt{\rho _{0}a^{-3(\omega +1)}\left( \alpha
_{0}(\omega -1)a^{3n}+2\right) }},\nonumber\\
\eea
where $\,_{2}F_{1}\left(a,b;c,z\right)$ is the hypergeometric function $\,_{2}F_{1}\left(a,b;c,z\right)=\sum _{k=0}^{\infty}{\left[(a)_k(b)_k/(c)_k\right]\left(z^k/k!\right)}$, and $t_0$ is an arbitrary constant of integration. For the deceleration parameter we obtain the expression
\be
q(a)=\frac{1}{2} \left\{n \left[\frac{6}{\alpha _0 (\omega -1) a^{3 n}+2}-3\right]+3 \omega +1\right\}.
\ee
In the limit of small values of the scale factor $\alpha _0 (\omega -1) a^{3 n}<<2$, we obtain $q\approx (1+3\omega)/2>0$, indicating a decelerating expansion during the early stages of the cosmological evolution. For $\alpha _0 (\omega -1) a^{3 n}>>2$, and for very large values of $a$, $q\approx \left[3(\omega -n)+1\right]/2=4(\omega +2)/(1-\omega )$.  Since generally for any realistic cosmological matter equation of state $\omega <1$, it follows that in both small and large time limits the time evolution of the Universe is decelerating.

\subsubsection{Non-conservative cosmological models with \(\nabla^\nu T_{\mu\nu}\neq 0\)}

By taking into account that \(T=\rho-3p\), from Eq.~(\ref{emtc}) we immediately obtain
\bea
-2(1+\alpha)\nabla^\nu T_{\mu\nu}&=&\nabla^\nu\alpha\frac{4T_{\mu\nu}-Tg_{\mu\nu}+2\theta_{\mu\nu}}{1-\alpha}+\nonumber\\
&&\alpha\left(2\nabla^\nu \theta_{\mu\nu}-g_{\mu\nu}T\right),
\eea
or, equivalently,
\begin{eqnarray}\label{35}
-2(1-\alpha)\nabla^\nu T_{\mu\nu}&=&\nabla^\nu\alpha\frac{g_{\mu\nu}\left(2L_m-T\right)}{1-\alpha}+\nonumber\\
&&\alpha g_{\mu\nu}\nabla^\nu \left(2L_m-T\right).
\end{eqnarray}
After multiplying Eq.~(\ref{35}) by the four-velocity vector $u^{\mu}$, defined in the comoving reference frame, we obtain
\begin{eqnarray}
u^\mu\nabla^\nu T_{\mu\nu}=\frac{1}{2(1-\alpha)}\left[\dot{\alpha}\frac{\rho-p}{1-\alpha}+\alpha(\dot{\rho}-\dot{p})\right].
\end{eqnarray}
By taking into account the mathematical identity
\be
u^\mu\nabla ^\nu T_{\mu\nu}=\dot{\rho}+3H(\rho+p)
\ee
we immediately find
\begin{eqnarray}\label{38}
\hspace{-0.7cm}\dot{\rho}+3H(\rho+p)&=&\frac{1}{2(1-\alpha)}\times \nonumber\\
&&\left[\dot{\alpha}(\rho-p)\left(\frac{1}{1-\alpha}\right)+\alpha(\dot{\rho}-\dot{p})\right].
\eea
For a general linear barotropic equation of state of the for $p=\omega (t)\rho$, Eq.~(\ref{38}) becomes
\bea
\left(1-\frac{3}{2}\alpha+\frac{1}{2}\omega\alpha\right)\dot{\rho}+\frac{1}{2}\alpha\rho\dot{\omega}
&=&3(\alpha-1)(1+\omega)H\rho+\nonumber\\
&&\frac{\dot{\alpha}(\omega-1)}{2(\alpha-1)}\rho.
\end{eqnarray}

For $\omega ={\rm constant}$ we have
\begin{eqnarray}\label{43}
\left(1-\frac{3}{2}\alpha+\frac{1}{2}\omega\alpha\right)\dot{\rho}
&=&3(\alpha-1)(1+\omega)H\rho+\nonumber\\
&&\frac{\dot{\alpha}}{2(\alpha-1)}(\omega-1)\rho .
\eea
After taking the derivative of Eq.~(\ref{f1c}) with respect to the time, and after substituting $\dot{H}$ with the use of Eq.~(\ref{f4c}), we obtain for the time deriative of the density the equation
\be\label{44}
\dot{\rho}=-\frac{6 H^2 \left[6 (\omega +1)(\alpha -1)^2 H +(\omega -1) \dot{\alpha }
   \right]}{\kappa ^2 \left[(\omega -3) \alpha +2\right]^2}.
\ee
Substituting Eq.~(\ref{44}) into Eq.~(\ref{43}) gives the equation
\bea\label{45}
&&\left[ 6(\omega +1)H(\alpha -1)^{2}+(\omega -1)\dot{\alpha}\right] \times \nonumber\\
&&\left\{
6H^{2}(\alpha -1)+\kappa ^{2}\rho \left[ (\omega -3)\alpha +2\right]
\right\} =0.
\eea
Due to the first generalized Friedmann equation (\ref{f1c}), Eq.~(\ref{45}) is identically satisfied during the cosmological evolution. However, a second solution of the field equation can be obtained by also imposing the condition that the first term in Eq.~(\ref{45}0 also vanishes identically.

\paragraph{The case $\alpha =1-\frac{1-\omega}{6(1+\omega)}\frac{1}{\ln\left( a/a_0\right)}$}

By requiring that the first term in the left-hand side of Eq.~(\ref{45}) also vanishes, we obtain for the scalar function $\alpha (t)$ the differential equation
\be
\dot{\alpha }= \frac{6 (1+\omega )  (1-\alpha )^2}{1-\omega}H,
\ee
from which we obtain
\be
\alpha (t)=1-\frac{1-\omega}{6(1+\omega)}\frac{1}{\ln \left[a(t)/a_0\right]},
\ee
where $a_0$ is an arbitrary constant of integration. By substituting this expression of $\alpha $ into Eq.~(\ref{dda}), we obtain the following second order differential equation describing the time evolution of the scale factor,
\be
\frac{\ddot{a}}{\dot{a}}=\left[1-\frac{3 (\omega +1)}{6 (\omega +1) \ln (Ca)+\omega -3}\right]\left(\frac{\dot{a}}{a}\right).
\ee
By integration we first obtain
\be
H=\frac{\dot{a}}{a}=  \frac{\zeta}{\sqrt{6 (\omega +1) \ln \left(a /a_0 \right)+\omega -3}},
\ee
where $\zeta$ is an arbitrary integration constant. Hence for the time variation of the scale factor we obtain
\be
a(t)=a_0e^{-\frac{\omega -3}{6(1+\omega)}}e^{\frac{3^{1/3}\zeta ^{2/3}}{2(1+\omega)^{1/3}}\left(t-t_0\right)^{2/3}},
\ee
where $t_0$ is an arbitrary constant of integration. For the time variation of the Hubble function we obtain
\be
H(t)=\frac{\zeta ^{2/3}}{3^{2/3} \sqrt[3]{\omega +1} \sqrt[3]{t-t_0}},
\ee
while the deceleration parameter $q$ of this model is given by
\be
q(t)=\frac{(1+\omega)^{1/3}}{3^{1/3} \zeta ^{2/3}
   \left(t-t_0\right)^{2/3}}-1.
\ee
In the limit of large times $t\rightarrow \infty $, we have $q\rightarrow -1$, and therefore the Universe ends in an (approximately de Sitter) accelerating phase. For the time variation of the function $\alpha $ we find
\be
\alpha (t)=\frac{1-\omega }{-3^{4/3} \zeta ^{2/3} (\omega +1)^{2/3}
   \left(t-t_0\right)^{2/3}+\omega -3}+1.
\ee
In the limit of large times $\alpha (t)$ tends to a constant, $\lim_{t\rightarrow \infty}\alpha (t)=1$.

\section{Modified gravity from quantum fluctuations proportional to the matter energy-momentum tensor--\(K^{\mu\nu}=\alpha T^{\mu\nu}\)}\label{sect4}

 Many extensions of standard general theory of relativity are based on the assumption that in certain physical situations space-time and matter may couple to each other \cite{fL1,fT1,Har4}. Hence it is natural to also consider the  case in which the average of the quantum fluctuations of the metric is proportional to the matter energy-momentum tensor,  \(K^{\mu\nu}=\alpha T^{\mu\nu}\), where \(\alpha\) is a constant. This approach suggests that the quantum perturbations of the space-time may also be strongly influenced  by the presence of the classical matter.

 \subsection{The gravitational field equations}

With the choice  \(K^{\mu\nu}=\alpha T^{\mu\nu}\) of the classical form of the average of the quantum fluctuations of the metric tensor we obtain for the first order quantum corrected gravitational Lagrangian the expression
\begin{eqnarray}\label{fT1}
\mathcal{L}&=&-\frac{1}{2k^2}\sqrt{-g}\left[R+\alpha \left(R_{\mu\nu}-\frac{1}{2}Rg_{\mu\nu}\right)T^{\mu\nu}\right]+\nonumber\\
&&\sqrt{-g}\left[L_m+\frac{1}{2}\alpha T_{\mu\nu}T^{\mu\nu}\right]=\nonumber\\
&&-\frac{1}{2k^2}\sqrt{-g}\left[R\left(1-\frac{1}{2}\alpha T\right)+\alpha R_{\mu\nu}T^{\mu\nu}\right]\nonumber\\
&&+\sqrt{-g}\left[\mathcal{L}_m+\frac{1}{2}\alpha T_{\mu\nu}T^{\mu\nu}\right].
\end{eqnarray}
By varying the gravitational action given by Eq.~(\ref{fT1}) with respect to the metric tensor $g^{\mu \nu}$ it follows that the classical gravitational field equations corresponding to the gravitational action (\ref{fT1}) are given by (for the full details of the derivation see Appendix~\ref{App2})
\begin{eqnarray}\label{ff1}
&&G_{\mu\nu}\left(1-\frac{1}{2}\alpha T\right)=\Bigg[\frac{1}{2}\alpha R\left(T_{\mu\nu}+\theta_{\mu\nu}\right)+\nonumber\\
&&\frac{1}{2}\alpha g_{\mu\nu}R_{\alpha\beta}T^{\alpha\beta}-\left(g_{\mu\nu}\Box-\nabla_\mu\nabla_\nu\right)\left(1-\frac{1}{2}\alpha T\right)\Bigg]-\nonumber\\
&&\kappa ^2\left[\frac{1}{2}\alpha g_{\mu\nu} T_{\alpha\beta}T^{\alpha\beta}-T_{\mu\nu}\right]+\nonumber\\
&&\kappa ^2\alpha\Bigg[T_{\alpha(\mu}T^\alpha_{\nu)}-T(g_{\mu\nu}\mathcal{L}_m+T_{\mu\nu})+2\mathcal{L}_m T_{\mu\nu}\Bigg]-\nonumber\\
&&\alpha\left[R_{\alpha(\mu}T^\alpha_{\nu)}-\frac{1}{2}R\left(g_{\mu\nu}\mathcal{L}_m+T_{\mu\nu}\right)+\mathcal{L}_m R_{\mu\nu}\right.+\nonumber\\
&&\left.\frac{1}{2}\left(\Box T_{\mu\nu}+\nabla_{\alpha}\nabla_{\beta}T^{\alpha\beta}g_{\mu\nu}-
\nabla_{\alpha}\nabla_{(\mu}T_{\nu)}^\alpha\right)\right],
\end{eqnarray}
where we have canceled the terms containing the second derivatives of \(\mathcal{L}_m\) with respect to the metric tensor, since  in most cases of physical interest they vanish. After contraction of Eq.~(\ref{ff1}) we find
\begin{eqnarray}
R\left(1-\frac{1}{2}\alpha T\right)-\alpha R\mathcal{L}_m+\alpha\left(\Box T-\nabla_\mu\nabla_\nu T^{\mu\nu}\right)\nonumber\\
+\kappa ^2(T-2\alpha T\mathcal{L}_m-\alpha T^2)=0.
\end{eqnarray}

Eq.~(\ref{ff1}) can then be rewritten as
\begin{eqnarray}
&&R_{\mu\nu}\left(1-\frac{1}{2}\alpha T+\alpha \mathcal{L}_m \right)+\alpha R_{\alpha(\mu}T_{\nu)}^\alpha- \frac{\alpha}{2} g_{\mu\nu}R_{\alpha\beta}T^{\alpha\beta}=\nonumber\\
&&-\frac{1}{2}g_{\mu\nu}\frac{1-\frac{1}{2}\alpha T}{1-\frac{1}{2}\alpha T-\alpha \mathcal{L}_m}\Bigg[\alpha\left(\Box T-\nabla_\alpha\nabla_\beta T^{\alpha\beta}\right)+\nonumber\\
&&2\kappa ^2\left(-\alpha TL_m-\frac{1}{2}\alpha T^2+\frac{1}{2}T\right)\Bigg]-\frac{1}{2}\alpha\times \nonumber\\
&&\Bigg(\Box T_{\mu\nu}+g_{\mu\nu}\nabla_\alpha\nabla_\beta T^{\alpha\beta}-\nabla_\alpha\nabla_{(\mu} T^{\alpha}_{\nu)}\Bigg)+\nonumber\\
&&\frac{1}{2}\alpha\left(g_{\mu\nu}\Box-\nabla_\mu\nabla\nu \right)T+2\kappa ^2 \times \nonumber\\
&&\left[\frac{1}{2}\alpha\left(T_{\alpha(\mu}T_{\nu)}^\alpha-g_{\mu\nu}T\mathcal{L}_m-T_{\mu\nu}T+2\mathcal{L}_mT_{\mu\nu}\right)\right.-\nonumber\\
&&\left.\frac{1}{4}\alpha g_{\mu\nu}T_{\alpha\beta}T^{\alpha\beta}+\frac{1}{2}T_{\mu\nu}\right].
\end{eqnarray}

\subsection{The divergence of the energy-momentum tensor}

By taking the divergence of the gravitational field equations (\ref{ff1})  we obtain first
\bea
\Big[\kappa^2\left(\alpha T-2\alpha \mathcal{L}_m-1\right)-\alpha R\Bigg]\nabla^\nu T_{\mu\nu}=\frac{1}{2}\alpha \times\nonumber\\
\Bigg[G_{\mu\nu} \nabla^\nu T+T_{\mu\nu}\nabla^\nu R+\nabla_\mu(R_{\alpha\beta}T^{\alpha\beta}-\kappa^2 T_{\alpha\beta}T^{\alpha\beta})\Bigg]\nonumber\\
+\kappa^2\alpha\Bigg[\nabla^\nu(T_{\alpha(\mu}T_{\nu)}^\alpha)-\nabla_\mu (T\mathcal{L}_m)-T_{\mu\nu}\times\nonumber\\
\nabla^\nu (T-2\mathcal{L}_m)\Bigg]-\alpha\Bigg[\nabla^\nu(R_{\alpha(\mu}T_{\nu)}^\alpha)-\frac{1}{2}\nabla_\mu(R\mathcal{L}_m)-\nonumber\\
\frac{1}{2}T_{\mu\nu}\nabla^\nu R+\nabla^\nu(R_{\mu\nu}\mathcal{L}_m)+\frac{1}{2}(\nabla^\nu\Box T_{\mu\nu}\nonumber\\
+\nabla_\mu\nabla_\alpha\nabla_\beta T^{\alpha\beta}-\nabla^\nu\nabla_\alpha\nabla_{(\mu}T_{\nu)}^\alpha)\Bigg].\nonumber\\
\eea

Hence for the divergence of the matter energy-momentum tensor in the modified gravity model induced by the quantum fluctuations of the metric proportional to the matter energy-momentum tensor we find
\bea
\nabla^\nu T_{\mu\nu}=\frac{1}{\kappa^2(\alpha T-2\alpha L_m-1)-\alpha R}\Bigg\{\frac{1}{2}\alpha \times\nonumber\\
\Bigg[T_{\mu\nu}\nabla^\nu R+\nabla_\mu(R_{\alpha\beta}T^{\alpha\beta}-\kappa^2 T_{\alpha\beta}T^{\alpha\beta})\Bigg]\nonumber\\
+\kappa^2\alpha\Bigg[\nabla^\nu(T_{\alpha(\mu}T_{\nu)}^\alpha)-\nabla_\mu (T\mathcal{L}_m)-T_{\mu\nu}\times\nonumber\\
\nabla^\nu (T-2\mathcal{L}_m)\Bigg]-\alpha\Bigg[\nabla^\nu(R_{\alpha(\mu}T_{\nu)}^\alpha)-\frac{1}{2}\nabla_\mu(R\mathcal{L}_m)-\nonumber\\
\frac{1}{2}T_{\mu\nu}\nabla^\nu R+\nabla^\nu(R_{\mu\nu}\mathcal{L}_m)+\frac{1}{2}(\nabla^\nu\Box T_{\mu\nu}\nonumber\\
+\nabla_\mu\nabla_\alpha\nabla_\beta T^{\alpha\beta}-\nabla^\nu\nabla_\alpha\nabla_{(\mu}T_{\nu)}^\alpha)\Bigg]\Bigg\}\nonumber\\
+\frac{1}{2}\alpha\frac{2\nabla^\nu T}{2-\alpha T}\Bigg\{\Bigg[\frac{1}{2}\alpha R(T_{\mu\nu}+\theta_{\mu\nu})\nonumber\\
+\frac{1}{2}\alpha g_{\mu\nu}R_{\alpha\beta}T^{\alpha\beta}-(g_{\mu\nu}\Box-\nabla_\mu\nabla_\nu)(1-\frac{1}{2}\alpha T)\Bigg]\nonumber\\
-\kappa ^2\left[\frac{1}{2}\alpha g_{\mu\nu} T_{\alpha\beta}T^{\alpha\beta}-T_{\mu\nu}\right]\nonumber\\
+\kappa ^2\alpha\Bigg[T_{\alpha(\mu}T^\alpha_{\nu)}-T(g_{\mu\nu}\mathcal{L}_m+T_{\mu\nu})+2\mathcal{L}_m T_{\mu\nu}\Bigg]\nonumber\\
-\alpha\left[R_{\alpha(\mu}T^\alpha_{\nu)}-\frac{1}{2}R(g_{\mu\nu}\mathcal{L}_m+T_{\mu\nu})+\mathcal{L}_m R_{\mu\nu}\right.\nonumber\\
\left.+\frac{1}{2}\left(\Box T_{\mu\nu}+\nabla_{\alpha}\nabla_{\beta}T^{\alpha\beta}g_{\mu\nu}-
\nabla_{\alpha}\nabla_{(\mu}T_{\nu)}^\alpha\right)\right]\Bigg\}. \nonumber\\
\eea
The above results show that generally in this class of models the matter energy-momentum tensor is not conserved. The non-conservation of $T_{\mu \nu}$ can be related to particle production processes that takes place due to the quantum fluctuations of the space-time metric.

\subsection{Cosmological applications}

For simplicity in the following we consider a spatially flat space-time, in which \(K=0\). By taking into account the intermediate results
\begin{eqnarray}
\Box T&=&(\ddot{\rho}-3\ddot{p})+(\dot{\rho}-3\dot{p})\frac{3\dot{a}}{a},\nonumber\\
\nabla^\alpha\nabla^\beta T_{\alpha\beta}&=&\frac{3\dot{a}}{a}\Bigg[(2\dot{\rho}+\dot{p})+2(\rho+p)\frac{\dot{a}}{a}\Bigg]+\nonumber\\
&&\ddot{\rho}+3(\rho+p)\frac{\ddot{a}}{a},
\end{eqnarray}
\begin{eqnarray}
\Box T_{\mu\nu}&=&\ddot{\rho}+\dot{\rho}\frac{3\dot{a}}{a},
\nabla_\mu \nabla_\nu T=\ddot{\rho}-3\ddot{p},
\nabla_\alpha\nabla_{(\mu} T^{\alpha}_{\nu)})\nonumber\\
&=&2\ddot{\rho}+2\dot{\rho}\frac{3\dot{a}}{a}, \mu=\nu=0,
\end{eqnarray}
we obtain first from the 00 component of Eq.~(\ref{ff1})
\begin{eqnarray}\label{111}
\frac{3\ddot{a}}{a}\Bigg[1+\alpha(\rho+3p)\Bigg]+\frac{\dot{a}^2}{a^2}3\alpha p=\nonumber\\
\frac{1}{2}\frac{1-\frac{1}{2}\alpha(\rho-3p)}{1-\frac{1}{2}\alpha(\rho- p)}\Bigg[-\alpha\frac{3\dot{a}}{a}\Bigg(\dot{\rho}+4\dot{p}+2(\rho+p)\frac{\dot{a}}{a}\Bigg)\nonumber\\
-3\alpha\ddot{p}-3\alpha(\rho+p)\frac{\ddot{a}}{a}+\kappa^2(\rho-3p)(1-\alpha(\rho-p))\Bigg]\nonumber\\
+\frac{\alpha}{2}\Bigg\{3(\rho+p)\frac{\ddot{a}}{a}+\frac{3\dot{a}}{a}\Bigg[4\dot{p}+2(\rho+p)\frac{\dot{a}}{a}\Bigg]\Bigg\}\nonumber\\
-\kappa^2\left[\rho-\alpha\left(\frac{1}{2}\rho^2+4\rho p+\frac{3}{2}p^2\right)\right].
\end{eqnarray}

For \(\mu=\nu=1\) we obtain
\begin{eqnarray}
\Box T_{\mu\nu}=a^2(\ddot{p}+\dot{p}\frac{3\dot{a}}{a}),
\nabla_\mu \nabla_\nu T=0,
\nabla_\alpha\nabla_{(\mu} T^{\alpha}_{\nu)})=0.\nonumber\\
\end{eqnarray}
Hence the 11 component of Eq.~(\ref{ff1}) gives
\begin{eqnarray}\label{112}
\frac{\dot{a}^2}{a^2}(2-\alpha(\rho-5p))+\frac{\ddot{a}}{a}(1-\alpha(2\rho-3p))+2\alpha\dot{p}\frac{\dot{a}}{a}=\nonumber\\
\frac{1}{2}\frac{1-\frac{1}{2}\alpha(\rho-3p)}{1-\frac{1}{2}\alpha(\rho- p)}\Bigg[-\alpha\frac{3\dot{a}}{a}\Bigg(\dot{\rho}+4\dot{p}+2(\rho+p)\frac{\dot{a}}{a}\Bigg)\nonumber\\
-3\alpha\ddot{p}-3\alpha(\rho+p)\frac{\ddot{a}}{a}+\kappa^2(\rho-3p)(1-\alpha\rho+\alpha p)\Bigg]\nonumber\\
+\frac{1}{2}\alpha\Bigg\{2\ddot{p}+3(\rho+p)\frac{\ddot{a}}{a}+\frac{3\dot{a}}{a}\Bigg[\dot{\rho}+3\dot{p}+2(\rho+p)\frac{\dot{a}}{a}\Bigg]\Bigg\}\nonumber\\
+\kappa^2\left(\frac{1}{2}\alpha(\rho^2+3p^2)+p\right).\nonumber\\
\end{eqnarray}

From Eqs.~(\ref{111}) and (\ref{112}) we obtain
\begin{eqnarray}
\frac{\ddot{a}}{a}=\frac{\Gamma \cdot \Delta-\Theta \cdot \Lambda}{\Xi},\\
\frac{\dot{a}^2}{a^2}=-\frac{\Pi\cdot \Gamma -\Sigma\cdot k}{\Xi}.\\
\end{eqnarray}
Therefore the generalized Friedmann equations for this gravity model take the form
\bea
3H^2=-3\frac{\Pi\cdot \Gamma-\Sigma\cdot \Theta}{\Xi},\\
2\dot{H}+3H^2=\frac{\Gamma\cdot (2\Delta-\Pi)+(\Sigma-l2)\cdot \Theta}{\Xi},
\eea
where we have denoted
\bea
\lambda=\frac{1-\frac{1}{2}\alpha(\rho-3p)}{1-\frac{1}{2}\alpha(\rho- p)},
\eea,
\bea
\Gamma=\frac{3}{2} \alpha  \lambda \left( \frac{\dot{a}}{a} \dot{\rho}+\ddot{p}\right)+6 \alpha  \frac{\dot{a}}{a} (\lambda-1) \dot{p}+\nonumber\\
\kappa^2 \left[\rho -\alpha  \left(\frac{3 p^2}{2}+4 p \rho +\frac{\rho ^2}{2}\right)\right]\nonumber\\
-\frac{1}{2} \lambda  \kappa^2 (\rho -3 p) \left(-\alpha  \rho +\alpha  p+1\right),
\eea
\bea
\Delta=2-\alpha ( \rho-5p) +3 \alpha  (\lambda -1) (p+\rho ),
\eea
\bea
\Theta=\frac{3}{2} \alpha (\lambda-1)  \frac{\dot{a}}{a}  \dot{\rho}+6 \alpha \frac{\dot{a}}{a} \lambda  \dot{p}-\frac{5 \alpha \dot{p}}{2}\frac{\dot{a}}{a} \nonumber\\
-\kappa^2\left(\frac{\alpha  \rho ^2}{2}+\frac{3 \alpha  p^2}{2}+p\right)+\frac{3 \alpha  \lambda  \ddot{p}}{2}-\nonumber\\
\alpha \ddot{p}-\frac{1}{2} \lambda \kappa^2 (\rho -3 p) (-\alpha  \rho +\alpha  p+1),
\eea
\bea
\Lambda=3 \alpha  \lambda ( \rho+p) -3 \alpha  \rho ,
\eea
\bea
\Xi=3\Bigg[\frac{\alpha ^2 \rho ^2}{2}(11-9\lambda)-3 \alpha  \lambda ( \rho+p)\nonumber\\
+ \alpha (2 \rho-7p) -(5+7\lambda ) \alpha ^2  p^2\nonumber\\
 +\frac{\alpha ^2 }{2}p \rho(15-23\lambda)-2\Bigg],
\eea
\bea
\Pi=\frac{3 \alpha  \lambda  }{2}(\rho +p)+\frac{ \alpha}{2}(3p-7\rho )+1,
\eea
\bea
\Sigma=3\Bigg[\frac{\alpha}{2}  ( \lambda-1)  (p+\rho )+\alpha  (3 p+\rho )+1\Bigg].
\eea
For the deceleration parameter we obtain the expression
\be
q=\frac{d}{dt}\frac{1}{H}-1=-\frac{\ddot{a}}{aH^2}=\frac{\Gamma\cdot \Delta-\Theta\cdot \Lambda}{\Pi\cdot \Gamma-\Sigma\cdot \Theta}.
\ee

For \(\alpha=0\) we have
$\Gamma=\kappa^2(\rho c^2+3p)/2$, $\Delta=2$, $\Theta=-\kappa^2(\rho c^2+p)/2$, $\Lambda=0$,
$\Xi=-6$, $\Pi=1$, and $\Sigma=3$, respectively. Hence in this limit we recover the standard Friedmann equations of general relativity.

\subsection{ Dust cosmological models with $p=0$}

By assuming that the matter content of the Universe consists of pressureless dust with $p=0$, we obtain immediately $\lambda=1$. Then the field equations can be written as
\bea
\frac{\ddot{a}}{a}=-\frac{3 \alpha H\dot{\rho}+\kappa^2 \rho }{6 (\alpha  \rho +1)},
\eea
and
\bea\label{128}
H^2=-\frac{3 \alpha  H \dot{\rho}(1-2\alpha\rho)+\kappa^2 \rho(4+\alpha\rho) }{6 (\alpha  \rho -2) (\alpha  \rho +1)},
\eea
respectively. Thus we obtain the generalized Friedmann equations of the present model as
\bea\label{129}
3H^2=-\frac{3 \alpha  H \dot{\rho}(1-2\alpha\rho)+\kappa^2 \rho(4+\alpha\rho) }{2 (\alpha  \rho -2) (\alpha  \rho +1)},
\eea
\bea\label{130}
2\dot{H}+3H^2=-\frac{\alpha  (\kappa^2 \rho ^2-3H\dot{\rho})}{2 (\alpha  \rho -2) (\alpha  \rho +1)}.
\eea
For the deceleration parameter we obtain
\bea
q=\frac{(3 \alpha H\dot{\rho}+k^2 \rho)(\alpha  \rho -2)}{3 \alpha  H \dot{\rho}(1-2\alpha\rho)+\kappa^2 \rho(4+\alpha\rho)}.
\eea

By multiplying with \(a^3\) both sides of Eq.~(129), taking its time derivative, and considering Eq.~(130), we obtain the time evolution of the matter density as
\bea
\ddot{\rho}H=-\dot{\rho}\frac{\ddot{a}}{a}+\frac{1}{-2\alpha+3\alpha^2\rho+3\alpha^3\rho^2-2\alpha^4\rho^3}\times\nonumber\\
\Bigg\{\dot{\rho}\Bigg[\frac{\kappa^2}{3}(8+4\alpha\rho+5\alpha^2\rho^2)+\alpha H^2(10-3\alpha\rho-\nonumber\\
9\alpha^2\rho^2+4\alpha^3\rho^3)-H\dot{\rho}(5\alpha^2-2\alpha^3\rho+2\alpha^4\rho^2)\Bigg]\nonumber\\
+\kappa^2H(8\rho+\alpha\rho^2-4\alpha^2\rho^3)\Bigg\}.
\eea

\subsubsection{de Sitter type evolution of the dust Universe}

By assuming that $H=H_0={\rm constant}$, the cosmological evolution equations (\ref{129}) and (\ref{130}) give for the time evolution of the density the first order differential equation
\be
\dot{\rho} (t)= \frac{2 \kappa ^2 \rho (t)}{3 \alpha  H_0 [\alpha
   \rho (t)-1]},
\ee
with the general solution given by
\be
\alpha  \rho -\ln (\rho )= \frac{2\kappa ^2}{3\alpha H_0}\left(t-t_0\right),
\ee
where $t_0$ is an arbitrary constant of integration. When $\alpha \rho >>\ln (\rho)$, the matter energy density linearly increases in time as $\rho (t)\propto t$, indicating that the late time de Sitter expansion of the Universe is triggered by (essentially quantum) particle creation processes.

\subsubsection{Cosmological evolution of the dust Universe}

From Eq.~(\ref{128}) we obtain for the time evolution of the matter density the differential equation
\be\label{135}
\dot{\rho }=\frac{6 H^2 (\alpha  \rho -2) (\alpha  \rho
   +1)+\kappa ^2 \rho  (\alpha  \rho +4)}{3 \alpha  H (2 \alpha  \rho
   -1)}.
\ee
Substitution of this equation into Eq.~(\ref{130}) gives the evolution equation of the Hubble function as
\be\label{136}
2\dot{H}=-3H^2+\frac{3 H^2-\kappa ^2 \rho }{2 \alpha  \rho -1}.
\ee
By introducing the set of dimensionless variables $\left(\tau, h, \alpha _0,r\right)$, defined as
\be
\tau =H_0t,H=H_0h,\alpha =\alpha _0\frac{\kappa ^2}{3H_0^2}, \rho =\frac{3H_0^2}{\kappa ^2}r,
\ee
where $H_0$ is the present day value of the Hubble function, and after changing the independent time variable from $\tau $ to the redshift $z$, Eqs.~(\ref{135}) and (\ref{136}) take the form
\be\label{137}
\frac{dr}{dz}=-\frac{2h^2\left(\alpha _0r-2\right)\left(\alpha _0r+1\right)+r\left(\alpha _0r+4\right)}{\alpha _0(1+z)h^2\left(2\alpha _0r-1\right)},
\ee
\be\label{138}
\frac{dh}{dz}=\frac{3}{2}\frac{h}{1+z}-\frac{3\left(h^2-r\right)}{2(1+z)h\left(2\alpha _0r-1\right)}.
\ee
The deceleration parameter is given by
\be
q=\frac{1}{2}-\frac{3\left(h^2-r\right)}{2h^2\left(2\alpha _0r-1\right)}.
\ee

The system of differential equation must be integrated with the initial conditions $h(0)=h_0$ and $r(0)=r_0$, respectively. It is interesting to note that for
$h(0)=r(0)=1$, the present day deceleration parameter takes the value $q(0)=1/2$. Thus, in order to obtain accelerating models we will adopt an initial condition for the matter energy density so that $r(0)=0.67$, while for $h$ we adopt the initial condition $h(0)=1$.  The variation with respect to the redshift $z$ of the dimensionless Hubble function $h$, of the dimensionless matter energy density $r$, and of the deceleration parameter $q$ are represented for $z\in [0,1]$ in Figs.~\ref{fig9} - \ref{fig11}, respectively.

\begin{figure}[tbp]
\includegraphics[width=7.5cm, angle=0]{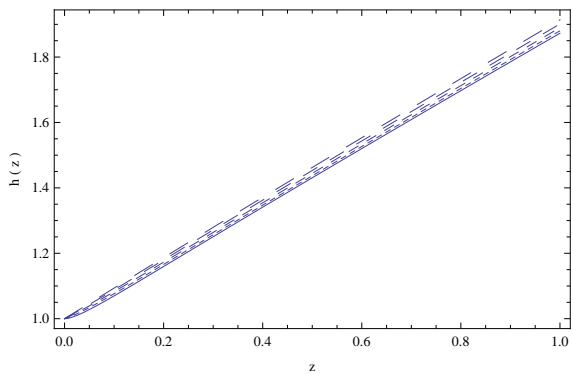} %
\caption{Variation with respect to the redshift of the dimensionless Hubble function $h(z)$ for  the modified gravity model with fluctuating quantum metric with $K_{\mu \nu} =\alpha T_{\mu \nu}$ for  $h(0)=1$, $r(0)=0.67$, and for different values of $\alpha _0$: $\alpha _0=1$ (solid curve), $\alpha _0=1.1$ (dotted curve), $\alpha _0=1.2$ (short dashed curve), $\alpha _0=1.3$ (dashed curve), and $\alpha _0=1.4$ (long dashed curve), respectively.}
\label{fig9}
\end{figure}

\begin{figure}[tbp]
\includegraphics[width=7.5cm, angle=0]{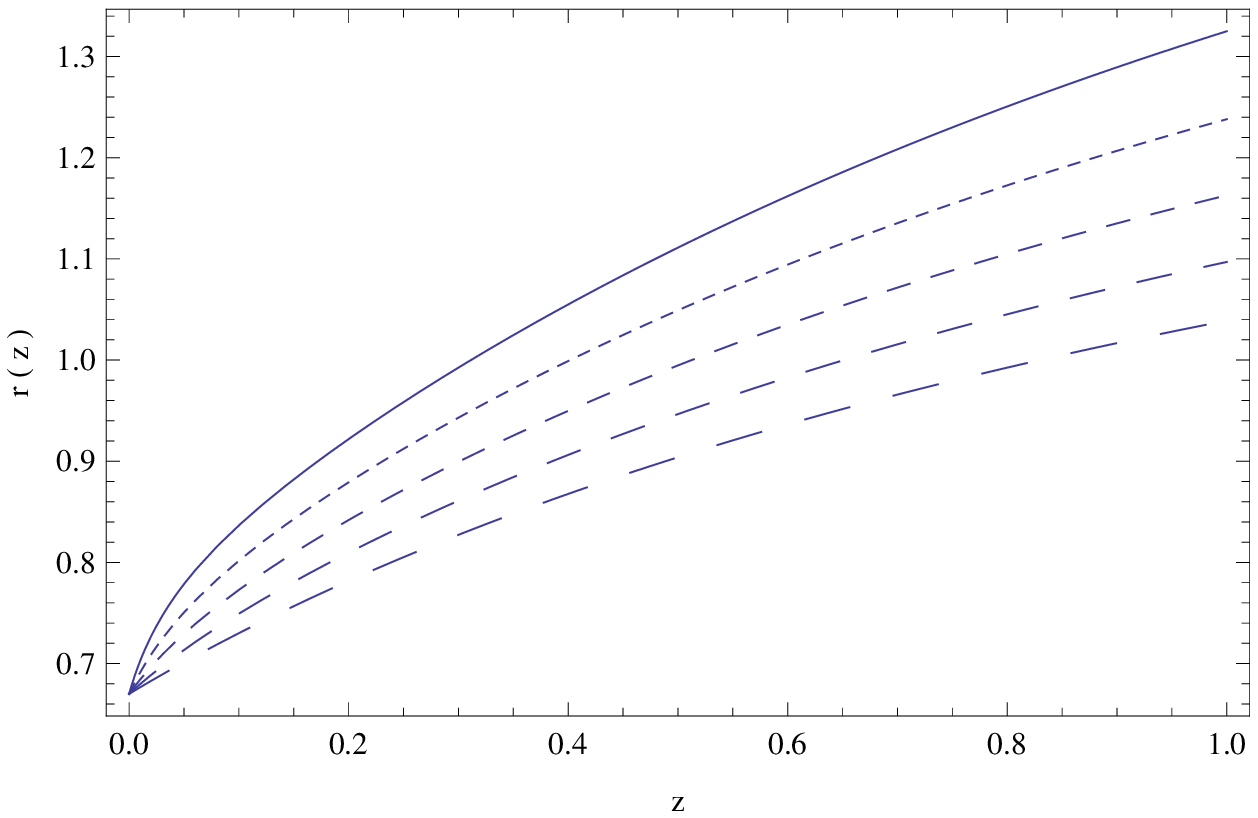} %
\caption{Variation with respect to the redshift of the dimensionless matter energy density $r(z)$ for the modified gravity model with fluctuating quantum metric with $K_{\mu \nu} =\alpha T_{\mu \nu}$ for  $h(0)=1$, $r(0)=0.67$, and for different values of $\alpha _0$: $\alpha _0=1$ (solid curve), $\alpha _0=1.1$ (dotted curve), $\alpha _0=1.2$ (short dashed curve), $\alpha _0=1.3$ (dashed curve), and $\alpha _0=1.4$ (long dashed curve), respectively.}
\label{fig10}
\end{figure}

\begin{figure}[tbp]
\includegraphics[width=7.5cm, angle=0]{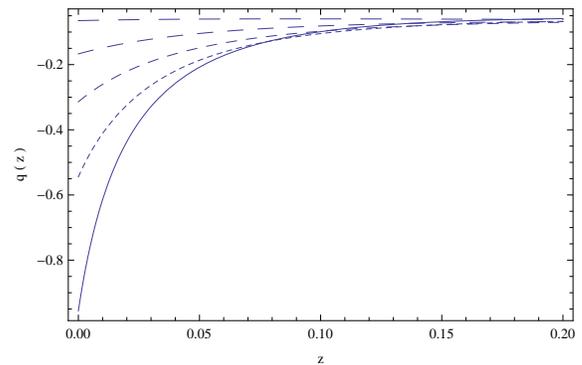} %
\caption{Variation with respect to the redshift of the deceleration parameter $q(z)$ for the modified gravity model with fluctuating quantum metric with $K_{\mu \nu} =\alpha T_{\mu \nu}$ for  $h(0)=1$, $r(0)=0.67$, and for different values of $\alpha _0$: $\alpha _0=1$ (solid curve), $\alpha _0=1.1$ (dotted curve), $\alpha _0=1.2$ (short dashed curve), $\alpha _0=1.3$ (dashed curve), and $\alpha _0=1.4$ (long dashed curve), respectively.}
\label{fig11}
\end{figure}

The Hubble function, depicted in Fig.~\ref{fig9}, is a monotonically increasing function of $z$, indication an expansionary evolution of the Universe. The numerical values of $h$ in the chosen redshift range show a very mild dependence on the  numerical values of the parameter $\alpha _0$. The energy density of the matter $r$, presented in Fig.~\ref{fig10}, is also a monotonically increasing function of the redshift, indicating a time decrease of $r$ during the cosmological evolution. The numerical values of $r$ depend strongly on $\alpha _0$. Finally, the deceleration parameter $q$, shown in Fig.~\ref{fig11}, also has a strong dependence on $\alpha _0$. In the range $z\in (0.14,1]$, the Universe is in a marginally decelerating phase, with $q\approx 0$. In this redshift range the evolution is independent on the numerical values of $\alpha _0$. For $z<0.14$, the Universe enters in an accelerating phase, and, depending on the values of $\alpha _0$, a large range of accelerating models can be constructed. The numerical values of $q$ rapidly increase with increasing $\alpha _0$, so that for $\alpha _0=1$, $q(0)\approx -1$, while for $\alpha _0=1.4$, the deceleration parameter $q$ is constant in all range $z\in [0,1]$, $q\approx 0$.

\section{Discussions and final remarks}\label{sect5}

In the present paper we have considered the cosmological properties of some classes of modified gravity models that are obtained from a first order correction of the quantum metric, as proposed in \cite{re8,re11}. By assuming that the quantum metric can be decomposed into two components, and by substituting the fluctuating part by its (classical) average value $K_{\mu \nu}$, a large class of modified gravity models can be obtained. As a first step in our study we have derived the general Einstein equations corresponding to an arbitrary $K_{\mu \nu}$. An important property of this class of models is the non-conservation of the matter energy-momentum tensor, which can be related to the physical processes of particle creation due to the quantum effects in the curved space-time. By assuming that $K_{\mu \nu}\propto g_{\mu \nu}$, a particular class of the modified $f(R,T)$ gravity theory is obtained. We have investigated in detail the cosmological properties of these models, by assuming that the coupling coefficient between the metric and the average value of the quantum fluctuation tensor is a scalar field with a non-vanishing self-interaction potential, and a simple scalar function. The scalar field self-interaction potential was assumed to be of Higgs type \cite{Aad}, which plays a fundamental role in elementary particle physics as describing the generation of mass in the quantum world.

We have investigated two cosmological models, in which the scalar field is in the minimum of the Higgs potential, and the case of the "complete" Higgs potential. In both approaches in the large time limit the Universe enters an accelerating phase, with the accelerating de Sitter solution acting as an attractor for these cosmological models. However, in the case of the "complete" Higgs potential the redshift evolution of the deceleration parameter $q$ indicates at low redshifts an extremely complex, oscillating behavior. Such a dynamics, as well as the corresponding cosmological evolution may play a significant role in the inflationary/post inflationary reheating phase in the history of the Universe. By assuming that the coupling between the metric and the quantum fluctuations is given by a scalar function, two distinct cases of cosmological models can be obtained. By imposing the conservativity of the energy-momentum tensor for a Universe filled with matter obeying a linear barotropic equation of state, a decelarating cosmological model is obtained. Thus a model could be useful to describe the evolution of the  high density Universe at large redshifts. An alternative model with matter creation can be also constructed, by imposing a specific equation of evolution for the time evolution of the coupling function $\alpha $. This model leads to an approximately de Sitter type expansion,  with the deceleration parameter tending to minus one in the large time limit,

A second modified gravity model can be obtained by assuming that the average value of the quantum fluctuation tensor is proportional to the matter energy-momentum tensor, $K_{\mu \nu}\propto T_{\mu \nu}$. This choice leads to an extension of the $f\left(R,T,R_{\mu \nu}T^{\mu \nu}\right)$ gravity theory \cite{Har4,Odin}, with the quantum corrected action including an extra term $T_{\mu \nu}T^{\mu \nu}$. Hence by considering the effects of the quantum fluctuations of the metric proportional to the matter energy-momentum tensor a particular case of a general $f\left(R,T,R_{\mu \nu}T^{\mu \nu},T_{\mu \nu}T^{\mu \nu}\right)$ modified gravity theory is obtained. The numerical analysis of the cosmological evolution equations for this model show that, depending on the numerical values of the coupling constant $\alpha $, for a Universe filled with dust matter a large variety of cosmological behaviors can be obtained at low redshifts, with the deceleration parameter varying between a constant (approximately) zero value on a large redshift range, and a de Sitter phase reached at $z=0$.

Quantum gravity represents the greatest challenge present day theoretical physics faces. Since no exact solutions for this problem are known, resorting to some approximate methods for studying quantum effects in gravity seems to be the best way to follow. A promising path may be represented by the inclusion of some tensor fluctuating terms in the metric, whose quantum mechanical origin can be well understood. Interestingly enough, such an approach leads to classical gravity models involving geometry-matter coupling, as well as non-conservative matter energy-momentum tensors, and, consequently, to particle creation processes. Hence even the study of the gravitational models with first order quantum corrections can lead to a better understanding of the physical foundations of the modified gravity models with geometry-matter coupling. In the present paper we have investigated some of the cosmological implications of these models, and we have developed some basic tools that could be used to further investigate the quantum effects in gravity,

\section*{Acknowledgments}

T. H. would like to thank the Yat Sen School of the Sun Yat - Sen University in Guangzhou, P. R. China, for the kind hospitality offered during the preparation of this work. S.-D. L. would like to thank to the Natural Science Funding of Guangdong Province for financial support (2016A030313313).

\appendix

\section{The derivation of the quantum corrected gravitational field equations for an arbitrary $K_{\mu \nu}$}\label{App1}

We start from the gravitational action given by Eq.~(\ref{2}),
\begin{eqnarray}
\mathcal{L}&=&-\frac{1}{2\kappa ^2}\sqrt{-g}\Bigg(R+G_{\mu\nu}\delta\hat{g}^{\mu\nu}\Bigg)+\sqrt{-g}\Bigg(\mathcal{L}_m+\nonumber\\
&&\frac{1}{2}T_{\mu\nu}\delta\hat{g}^{\mu\nu}\Bigg).
\end{eqnarray}
After substituting the explicit form of the quantum metric fluctuation as  $\left<\delta \hat{g}^{\mu \nu}\right>=K^{\mu \nu}$, the gravitational Lagrangian becomes
\bea\label{A1}
\mathcal{L}&=&-\frac{1}{2\kappa ^2}\sqrt{-g}\Bigg(R+G_{\mu \nu}K^{\mu \nu}\Bigg)+\nonumber\\
&&\sqrt{-g}\Bigg(\mathcal{L}_m+\frac{1}{2}T_{\mu \nu}{K}^{\mu\nu}\Bigg).
\eea

By taking the variation with respect to the metric tensor we obtain first
\bea
\hspace{-0.5cm}\delta\cL&=&\sqrt{-g}\delta g^{\mu\nu}\left(\frac{1}{2}T_{\mu\nu}-\frac{1}{2\kappa^2}G_{\mu\nu}\right)-\nn
\hspace{-0.5cm}&&\frac{1}{2\kappa^2}\delta\left(\sqrt{-g}G_{\mu\nu}K^{\mu\nu}\right)+\frac{1}{2}\delta \left(\sqrt{-g}T_{\mu\nu}K^{\mu\nu}\right).
\eea
Then we obtain
\bea
\delta\left(\sqrt{-g}G_{\mu\nu}K^{\mu\nu}\right)=-\sqrt{-g}\delta g^{\mu\nu}\frac{1}{2}g_{\mu\nu}G_{\alpha\beta}K^{\alpha\beta}+\nn
\sqrt{-g}\Bigg[\delta\left(R_{\mu\nu}K^{\mu\nu}\right)-\frac{1}{2}\delta(RK)\Bigg]=\nn
-\sqrt{-g}\delta g^{\mu\nu}\frac{1}{2}g_{\mu\nu}G_{\alpha\beta}K^{\alpha\beta}+\sqrt{-g}\delta g^{\mu\nu}\frac{1}{2}\times\nn
\Bigg(\Box K_{\mu\nu}+\nabla_{\alpha}\nabla_{\beta}K^{\alpha\beta}g_{\mu\nu}-
\nabla_{\alpha}\nabla_{(\mu}K_{\nu)}^\alpha\Bigg)+\nn
\sqrt{-g}\delta g^{\mu\nu}(\gamma^{\alpha\beta}_{\mu\nu} R_{\alpha\beta})-\frac{1}{2}\sqrt{-g}\delta g^{\mu\nu}\Bigg(RK_{\mu\nu}+\nn
KR_{\mu\nu}+g_{\alpha\beta}\left(\gamma^{\alpha\beta}_{\mu\nu}R\right)+\nabla_\mu\nabla_\nu K+g_{\mu\nu}\Box K\Bigg),
\eea
 where we have defined \(K=g_{\mu\nu}K^{\mu\nu}\) and \(A_{\alpha\beta}\delta K^{\alpha\beta}=\delta g^{\mu\nu}(\gamma^{\alpha\beta}_{\mu\nu}A_{\alpha\beta})\). Here \(A_{\alpha\beta}=R_{\alpha \beta}\), or \(A_{\alpha\beta}=T_{\alpha \beta}\), and \(\gamma^{\alpha\beta}_{\mu\nu}\) can be a tensor, an operator, or the combination of them. Moreover,
\bea
\delta\left(\sqrt{-g}T_{\mu\nu}K^{\mu\nu}\right)=-\sqrt{-g}\delta g^{\mu\nu}\frac{1}{2}g_{\mu\nu}T_{\alpha\beta}K^{\alpha\beta}+\nn
\sqrt{-g}\delta g^{\mu\nu}\Bigg[\gamma^{\alpha\beta}_{\mu\nu}T_{\alpha\beta}+K^{\alpha\beta}(2\frac{\delta^2 L_m}{\delta g^{\mu\nu}\delta g^{\alpha\beta}}-\frac{1}{2}g_{\alpha\beta}T_{\mu\nu}\nn
-\frac{1}{2}g_{\mu\nu}g_{\alpha\beta}L_m-L_m\frac{\delta{g_{\alpha\beta}}}{\delta g^{\mu\nu}})\Bigg].
\eea
Therefore we finally obtain for the variation of the first order quantum corrected gravitational Lagrangian the expression
\bea
\delta\cL=\sqrt{-g}\delta g^{\mu\nu}\left(\frac{1}{2}T_{\mu\nu}-\frac{1}{2\kappa^2}G_{\mu\nu}\right)-\nn
\frac{1}{2\kappa^2}\Bigg\{-\sqrt{-g}\delta g^{\mu\nu}\frac{1}{2}g_{\mu\nu}G_{\alpha\beta}K^{\alpha\beta}+\sqrt{-g}\delta g^{\mu\nu}\frac{1}{2}\times\nn
\Bigg(\Box K_{\mu\nu}+\nabla_{\alpha}\nabla_{\beta}K^{\alpha\beta}g_{\mu\nu}-
\nabla_{\alpha}\nabla_{(\mu}K_{\nu)}^\alpha\Bigg)+\nn
\sqrt{-g}\delta g^{\mu\nu}(\gamma^{\alpha\beta}_{\mu\nu}R_{\alpha\beta})-\frac{1}{2}\sqrt{-g}\delta g^{\mu\nu}(RK_{\mu\nu}+\nn
KR_{\mu\nu}+\gamma^{\alpha\beta}_{\mu\nu}(R g_{\alpha\beta})+\nabla_\mu\nabla_\nu K+g_{\mu\nu}\Box K)\Bigg\}+\nn
\frac{1}{2}\Bigg\{-\sqrt{-g}\delta g^{\mu\nu}\frac{1}{2}g_{\mu\nu}T_{\alpha\beta}K^{\alpha\beta}+\nn
\sqrt{-g}\delta g^{\mu\nu}\Bigg[\gamma T_{\mu\nu}+K^{\alpha\beta}(2\frac{\delta^2 L_m}{\delta g^{\mu\nu}\delta g^{\alpha\beta}}-\frac{1}{2}g_{\alpha\beta}T_{\mu\nu}\nn
-\frac{1}{2}g_{\mu\nu}g_{\alpha\beta}L_m-L_m\frac{\delta{g_{\alpha\beta}}}{\delta g^{\mu\nu}})\Bigg]\Bigg\}.
\eea

Hence the gravitational field equations corresponding to the Lagrangian (\ref{A1}) now reads
\bea
G_{\mu\nu}=\kappa^2T_{\mu\nu}-
\Bigg\{\frac{1}{2}g_{\mu\nu}G_{\alpha\beta}K^{\alpha\beta}+\frac{1}{2}\times\nn
\Bigg(\Box K_{\mu\nu}+\nabla_{\alpha}\nabla_{\beta}K^{\alpha\beta}g_{\mu\nu}-
\nabla_{\alpha}\nabla_{(\mu}K_{\nu)}^\alpha\Bigg)+\nn
\gamma^{\alpha\beta}_{\mu\nu}R_{\alpha\beta}-\frac{1}{2}\Bigg[RK_{\mu\nu}+KR_{\mu\nu}+\gamma^{\alpha\beta}_{\mu\nu}(R g_{\alpha\beta})+\nn
\nabla_\mu\nabla_\nu K+g_{\mu\nu}\Box K\Bigg]\Bigg\}+\kappa^2\Bigg\{-\frac{1}{2}g_{\mu\nu}T_{\alpha\beta}K^{\alpha\beta}+\nn
\Bigg[\gamma^{\alpha\beta}_{\mu\nu}T_{\alpha\beta}+K^{\alpha\beta}(2\frac{\delta^2 L_m}{\delta g^{\mu\nu}\delta g^{\alpha\beta}}-\frac{1}{2}g_{\alpha\beta}T_{\mu\nu}\nn
-\frac{1}{2}g_{\mu\nu}g_{\alpha\beta}L_m-L_m\frac{\delta{g_{\alpha\beta}}}{\delta g^{\mu\nu}})\Bigg]\Bigg\}.
\eea

\section{The derivation of the field equations for $K_{\mu \nu}=\alpha T_{\mu \nu}$}\label{App2}

In order to obtain the field equations of the $K_{\mu \nu}=\alpha T_{\mu \nu}$ modified gravity model we begin by varying the gravitational action (\ref{fT1}) with respect to the metric tensor. Thus we obtain
\begin{eqnarray}
\delta\mathcal{L}&=&-\frac{1}{2k^2}\Bigg[\sqrt{-g}G_{\mu\nu}\left(1-\frac{1}{2}\alpha T\right)\delta g^{\mu\nu}-\frac{1}{2}\alpha\sqrt{-g}R\times\nonumber\\
&&\left(T_{\mu\nu}+\theta_{\mu\nu}\right)\delta g^{\mu\nu}+\alpha\delta(\sqrt{-g}R_{\mu\nu}T^{\mu\nu})+\nonumber\\
&&\sqrt{-g}\left(1-\frac{1}{2}\alpha T\right)(g_{\mu\nu}\Box-\nabla_\mu\nabla_\nu)\delta g^{\mu\nu}\Bigg]+\nonumber\\
&&\left[\frac{1}{2}\sqrt{-g}T_{\mu\nu}\delta g^{\mu\nu}+\frac{1}{2}\alpha\delta(\sqrt{-g}T_{\mu\nu}T^{\mu\nu})\right]=\nonumber\\
&&-\frac{1}{2k^2}\Bigg[\sqrt{-g}G_{\mu\nu}\left(1-\frac{1}{2}\alpha T\right)\delta g^{\mu\nu}-\frac{1}{2}\alpha\sqrt{-g}R\times\nonumber\\
&&(T_{\mu\nu}+\theta_{\mu\nu})\delta g^{\mu\nu}-\frac{1}{2}\alpha g_{\mu\nu}\sqrt{-g} R_{\alpha\beta}T^{\alpha\beta}\delta g^{\mu\nu}+\nonumber\\
&&\sqrt{-g}\delta g^{\mu\nu}\left(g_{\mu\nu}\Box-\nabla_\mu\nabla_\nu\right)\left(1-\frac{1}{2}\alpha T\right)\Bigg]-\nonumber\\
&&\left[\frac{1}{4}\alpha g_{\mu\nu}\sqrt{-g} T_{\alpha\beta}T^{\alpha\beta}\delta g^{\mu\nu}-\frac{1}{2}\sqrt{-g}T_{\mu\nu}\delta g^{\mu\nu}\right]+\nonumber\\
&&\frac{1}{2}\alpha\sqrt{-g}\delta\left(T_{\mu\nu}T^{\mu\nu}\right)-\frac{\alpha}{2k^2}\sqrt{-g}\delta \left(R_{\mu\nu}T^{\mu\nu}\right).
\end{eqnarray}

We shall now calculate the variation of the terms \(\delta(T_{\mu\nu}T^{\mu\nu})\) and \(\delta (R_{\mu\nu}T^{\mu\nu})\), respectively. In order to do this we take into account the following known results,
\begin{eqnarray}
&\delta \Gamma _{\mu \nu }^{\lambda }=\frac{1}{2}g^{\lambda \sigma }\left(
\nabla _{\nu }\delta g_{\mu \sigma }+\nabla _{\mu }\delta g_{\nu \sigma
}-\nabla _{\sigma }\delta g_{\mu \nu }\right), \nonumber\\
&\nabla _{\lambda}\delta \Gamma _{\mu \nu }^{\lambda }=\frac{1}{2}\left(
\nabla ^{\sigma}\nabla _{\nu }\delta g_{\mu \sigma }+\nabla ^{\sigma}\nabla _{\mu }\delta g_{\nu \sigma
}-\nabla ^{\sigma}\nabla _{\sigma }\delta g_{\mu \nu }\right),\nonumber\\
&\delta \Gamma _{\mu \lambda }^{\lambda }=\frac{1}{2}g^{\lambda \sigma }\left(
\nabla _{\lambda }\delta g_{\mu \sigma }+\nabla _{\mu }\delta g_{\lambda \sigma
}-\nabla _{\sigma }\delta g_{\mu \lambda }\right),\nonumber\\
&\nabla _{\nu}\delta \Gamma _{\mu \lambda }^{\lambda }=\frac{1}{2}\left(
\nabla _{\nu}\nabla ^{\sigma}\delta g_{\mu \sigma }+g^{\lambda \sigma }\nabla _{\nu}\nabla _{\mu }\delta g_{\lambda \sigma}-\nabla _{\nu}\nabla ^{\lambda}\delta g_{\mu \lambda }\right).\nonumber
\end{eqnarray}

Hence
\begin{eqnarray}
&&T^{\mu\nu}\delta R_{\mu\nu}=\frac{1}{2}T^{\mu\nu}\Bigg(
\nabla ^{\sigma}\nabla _{\nu }\delta g_{\mu \sigma }+\nabla ^{\sigma}\nabla _{\mu }\delta g_{\nu \sigma
}-\nonumber\\
&&\nabla ^{\sigma}\nabla _{\sigma }\delta g_{\mu \nu }-g^{\lambda \sigma }\nabla _{\nu}\nabla _{\mu }\delta g_{\lambda \sigma}\Bigg)=\nonumber\\
&&\frac{\delta g^{\mu\nu}}{2}\left(\Box T_{\mu\nu}+\nabla_{\alpha}\nabla_{\beta}T^{\alpha\beta}g_{\mu\nu}-
\nabla_{\alpha}\nabla_{(\mu}T_{\nu)}^\alpha\right).
\end{eqnarray}
For the variation of the terms $R_{\mu\nu}\delta T^{\mu\nu}$ and $\delta R_{\mu\nu}T^{\mu\nu}$ we obtain
\begin{eqnarray}
R_{\mu\nu}\delta T^{\mu\nu}&=&R_{\mu\nu}\delta(g^{\mu\alpha}g^{\nu\beta}T_{\alpha\beta})
=R^{\alpha\beta}\delta T_{\alpha\beta}+R_{\alpha(\nu}T^\alpha_{\mu)}\delta g^{\mu\nu}\nonumber\\
&=&R_{\alpha(\nu}T^\alpha_{\mu)}\delta g^{\mu\nu}+\delta g^{\mu\nu}\Bigg[2R^{\alpha\beta}\frac{\partial^2 \mathcal{L}_m}{\partial g^{\mu\nu}\partial g^{\alpha\beta}}\nonumber\\
&&-\frac{1}{2}R(g_{\mu\nu}\mathcal{L}_m+T_{\mu\nu})-\mathcal{L}_m\frac{R^{\alpha\beta}\delta g_{\alpha\beta}}{\delta g^{\mu\nu}}\Bigg]\nonumber\\
&=&\delta g^{\mu\nu}\Bigg[R_{\alpha(\mu}T^\alpha_{\nu)}-\frac{1}{2}R(g_{\mu\nu}\mathcal{L}_m+T_{\mu\nu})\nonumber\\
&&+\mathcal{L}_m R_{\mu\nu}+2R^{\alpha\beta}\frac{\partial^2 \mathcal{L}_m}{\partial g^{\mu\nu}\partial g^{\alpha\beta}}\Bigg],
\end{eqnarray}
\begin{eqnarray}
&&\delta R_{\mu\nu}T^{\mu\nu}=\delta g^{\mu\nu}\Bigg[R_{\alpha(\mu}T^\alpha_{\nu)}-\frac{1}{2}R\Bigg(g_{\mu\nu}\mathcal{L}_m\nonumber\\
&&+T_{\mu\nu}\Bigg)+\mathcal{L}_m R_{\mu\nu}+2R^{\alpha\beta}\frac{\partial^2 \mathcal{L}_m}{\partial g^{\mu\nu}\partial g^{\alpha\beta}}\nonumber\\
&&+\frac{1}{2}\Bigg(\Box T_{\mu\nu}+\nabla_{\alpha}\nabla_{\beta}T^{\alpha\beta}g_{\mu\nu}-
\nabla_{\alpha}\nabla_{(\mu}T_{\nu)}^\alpha\Bigg)\Bigg].
\end{eqnarray}

Similarly,
\begin{eqnarray}
T_{\mu\nu}\delta T^{\mu\nu}=\delta g^{\mu\nu}\Bigg[T_{\alpha(\mu}T^\alpha_{\nu)}-\frac{1}{2}T\left(g_{\mu\nu}\mathcal{L}_m+T_{\mu\nu}\right)\nonumber\\
+\mathcal{L}_m T_{\mu\nu}+2T^{\alpha\beta}\frac{\partial^2 \mathcal{L}_m}{\partial g^{\mu\nu}\partial g^{\alpha\beta}}\Bigg],
\end{eqnarray}
\begin{eqnarray}
T^{\mu\nu}\delta T_{\mu\nu}=\delta g^{\mu\nu}\Bigg[-\frac{1}{2}T(g_{\mu\nu}\mathcal{L}_m+T_{\mu\nu})+L_m T_{\mu\nu}\nonumber\\
+2T^{\alpha\beta}\frac{\partial^2 \mathcal{L}_m}{\partial g^{\mu\nu}\partial g^{\alpha\beta}}\Bigg],
\end{eqnarray}
\begin{eqnarray}
\delta T_{\mu\nu}T^{\mu\nu}=\delta g^{\mu\nu}\Bigg[T_{\alpha(\mu}T^\alpha_{\nu)}-T\left(g_{\mu\nu}L_m+T_{\mu\nu}\right)+\nonumber\\
2\mathcal{L}_m T_{\mu\nu}+4T^{\alpha\beta}\frac{\partial^2 \mathcal{L}_m}{\partial g^{\mu\nu}\partial g^{\alpha\beta}}\Bigg].
\end{eqnarray}
And thus for the total variation of the action we obtain
\begin{eqnarray}
&&\delta\mathcal{L}
=-\frac{1}{2\kappa ^2}\Bigg[\sqrt{-g}G_{\mu\nu}\left(1-\frac{1}{2}\alpha T\right)\delta g^{\mu\nu}-\nonumber\\
&&\frac{\alpha\sqrt{-g}}{2}\Bigg[R(T_{\mu\nu}+\theta_{\mu\nu})+g_{\mu\nu}R_{\alpha\beta}T^{\alpha\beta}\Bigg]\delta g^{\mu\nu}+\nonumber\\
&&\sqrt{-g}\delta g^{\mu\nu}\left(g_{\mu\nu}\Box-\nabla_\mu\nabla_\nu\right)\left(1-\frac{1}{2}\alpha T\right)\Bigg]-\nonumber\\
&&\delta g^{\mu\nu}\sqrt{-g}\left[\frac{1}{4}\alpha g_{\mu\nu} T_{\alpha\beta}T^{\alpha\beta}-\frac{1}{2}T_{\mu\nu}\right]+\nonumber\\
&&\frac{\alpha\sqrt{-g}}{2}\delta g^{\mu\nu}\Bigg[T_{\alpha(\mu}T^\alpha_{\nu)}-T\left(g_{\mu\nu}L_m+T_{\mu\nu}\right)+\nonumber\\
&&2\mathcal{L}_m T_{\mu\nu}+4T^{\alpha\beta}\frac{\partial^2 \mathcal{L}_m}{\partial g^{\mu\nu}\partial g^{\alpha\beta}}\Bigg]-\nonumber\\
&&\frac{\alpha\sqrt{-g}}{2\kappa ^2}\delta g^{\mu\nu}\Bigg[R_{\alpha(\mu}T^\alpha_{\nu)}-\frac{1}{2}R\left(g_{\mu\nu}\mathcal{L}_m+T_{\mu\nu}\right)+\nonumber\\
&&\mathcal{L}_m R_{\mu\nu}+2R^{\alpha\beta}\frac{\partial^2 \mathcal{L}_m}{\partial g^{\mu\nu}\partial g^{\alpha\beta}}+\nonumber\\
&&\frac{1}{2}\Bigg(\Box T_{\mu\nu}+\nabla_{\alpha}\nabla_{\beta}T^{\alpha\beta}g_{\mu\nu}-
\nabla_{\alpha}\nabla_{(\mu}T_{\nu)}^\alpha\Bigg)\Bigg].
\end{eqnarray}

The condition $\delta \mathcal{L}=0$ gives immediately the field equations (\ref{ff1}).

\end{document}